\documentclass[a4paper,fleqn,usenatbib]{mnras}
\usepackage[T1]{fontenc}
\usepackage{ae,aecompl}
\usepackage{amsmath}
\usepackage{amsfonts}
\usepackage{amssymb}
\usepackage{graphicx}
\def\mnras{MNRAS}
\def\aap{A$\&$A}
\def\nat{Nature}
\def\apj{ApJ}
\def\apjl{ApJ Letters}
\def\apjs{ApJS}

\def\aj{AJ}
\def\icarus{Icarus}

\newcommand{\me}{{\, {\rm M}_{\oplus}}}
\newcommand{\msun}{{\, {\rm M}_{\odot}}}
\newcommand{\au}{{\, {\rm au}}}
\newcommand{\rgeuv}{{\, {r_{\rm g, euv}}}}
\newcommand{\rgfuv}{{\, {r_{\rm g, fuv}}}}
\newcommand{\Stokes}{{\, {\rm St}}}
\newcommand{\emb}{{{proto-embryo~}}}
\newcommand{\embs}{{{proto-embryos~}}}
\title[Embryo Formation]{From Dust to Planets I: Planetesimal and Embryo Formation}
\author[Coleman, G. A. L]{Gavin A. L. Coleman\thanks{Email: gavin.coleman@qmul.ac.uk}\\
Queen Mary University of London, Mile End Road, London, United Kingdom}
\date{Accepted 2021 July 1; Received 2021 June 29; in original form 2020 July 1}
\pubyear{2021}
\begin{document}
\label{firstpage}
\pagerange{\pageref{firstpage}--\pageref{lastpage}}
\maketitle
\begin{abstract}
Planet formation models begin with \embs and planetesimals already fully formed, missing out a crucial step, the formation of planetesimals/\embs.
In this work, we include prescriptions for planetesimal and \emb formation arising from pebbles becoming trapped in short-lived pressure bumps, in thermally evolving viscous discs to examine the sizes and distributions of \embs and planetesimals throughout the disc.
We find that planetesimal sizes increase with orbital distance, from $\sim$10 km close to the star to hundreds of kilometres further away.
Proto-embryo masses are also found to increase with orbital radius, ranging from $10^{-6}\me$ around the iceline, to $10^{-3}\me$ near the orbit of Pluto.
We include prescriptions for pebble and planetesimal accretion to examine the masses that \embs can attain.
Close to the star, planetesimal accretion is efficient due to small planetesimals, whilst pebble accretion is efficient where pebble sizes are fragmentation limited, but inefficient when drift dominated due to low accretion rates before the pebble supply diminishes.
Exterior to the iceline, planetesimal accretion becomes inefficient due to increasing planetesimal eccentricities, whilst pebble accretion becomes more efficient as the initial \emb masses increase, allowing them to significantly grow before the pebble supply is depleted.
Combining both scenarios allows for more massive \embs at larger distances, since the accretion of planetesimals allows pebble accretion to become more efficient, allowing giant planet cores to form at distances upto $10\au$.
By including more realistic initial \emb and planetesimal sizes, as well as combined accretion scenarios, should allow for a more complete understanding in the beginning to end process of how planets and planetary systems form.

\end{abstract}

\section{Introduction}
\label{sec:intro}

With the number of exoplanets discovered now exceeding 4000 \citep[e.g.][]{Winn15}, understanding their formation is one of the key problems in astrophysics.
Not only should planet formation models be able to reproduce the occurrence rates of planets but also the diversity.
This diversity, ranging from terrestrial and super-Earth mass planets in and out of resonance \citep{Anglada2016,Damasso20,GillonTrappist17} to hot and cold Jupiters \citep{MayorQueloz,Robertson12} would also need to be explained in these models.

Traditionally within the core accretion model, it was thought that planets formed through the accretion of planetesimals \citep{Pollack}, however this method of accretion was an issue for forming giant planets since the time-scales for formation, were typically longer than observed disc lifetimes \citep{Haisch01,Mamajek09,Ribas14}.
This was especially a problem when using 100 km planetesimals, thought to be the original planetesimal size based on planetesimal formation models \citep{Youdin05,Johansen07,JohansenYoudin2009,Bai10} as well as observations of the Solar System \citep{Morbidelli09,Delbo17}.
The way that many authors got around this problem was with the use of smaller planetesimals that allowed for much more efficient accretion rates \citep[e.g.][]{Alibert2006,Ida13,Mordasini15,ColemanNelson14,ColemanNelson16,ColemanNelson16b}.
Whilst smaller planetesimals gave more favourable accretion time-scales, the question always remained as to how the planetesimals and protoplanets formed with their initial sizes, given that theory and observations point towards larger planetesimal sizes.

More recently, in response to the time-scale issues of planetesimal accretion, a different accretion regime appeared.
This new regime involved planetary cores accreting pebbles as they drifted past the planet \citep{OrmelKlahr2010,Lambrechts12}.
Since pebbles are much smaller in size, they could then be easily accreted by the planetary cores, resulting in much faster formation time-scales.
This then led to numerous works that aimed to explain the observed exoplanet populations and their formation pathways \citep{Bitsch15,Lambrechts19,Bitsch19}.

Another question remains is as to how/whether pebble and planetesimal accretion scenarios can work in tandem within global disc models.
Recently numerous papers have directly compared pebble and planetesimal accretion scenarios in terms of the types of planets and planetary systems that arise from each.
\citet{ColemanProxima17} and \citet{Coleman19} found that both scenarios formed remarkably similar planetary systems around low mass stars such as Proxima Centauri and TRAPPIST-1.
These similarities included, planetary masses and periods, resonances between neighbouring planets, and general observability of the systems.
More recently \citet{Brugger20} examined the outcomes of single planet populations around Solar mass stars, finding that the planetesimal accretion scenario forms more giant planets, whilst the pebble accretion scenario forms more super-Earths.
This was found to be due to gas accretion being inhibited before the planets reached the pebble isolation mass, as a result of high solid accretion rates limiting the amount of gas that could be accreted.
This stopped the planets from being able to undergo runaway gas accretion, as because they grew early in the disc lifetime, migration forces were considerably stronger resulting in the planets migrating into the inner disc near to the central star before substantial gas envelopes were accreted.
The planetesimal accretion scenario did not have this problem since the weaker solid accretion rate allowed the planets to grow more slowly, resulting in weaker migration forces when migration becomes a more dominant process for those planets evolutions. The weaker solid accretion also allowed for a larger envelope to be retained by the planet whilst it was still low in mass, which then allowed the planet to accrete significant amounts of gas later in the disc lifetime, and undergo runaway gas accretion before fully migrating near to the central star.

Whilst the differences between the two scenarios have been compared, the similarity that these works contain, is that the initial embryos that become planets are typically much larger than what is thought to form through gravitational collapse following for example the streaming instability \citep{Johansen07}.
Planetesimal accretion models, typically use embryos of at least a Lunar mass ($0.0123\me$), whilst the pebble accretion models typically begin with embryos at the transition mass, which increases with orbital distance and typically ranges from $10^{-3} \me$ to $10^{-1} \me$.
In all of these cases, the initial embryos are much more massive than the characteristic size of the planetesimals that form through gravitational collapse, approximately a few hundred kilometres \citep{Johansen12,Johansen15,Simon16,Schafer17,Simon17,Abod19}.
Planetesimal accretion models also assume the planetesimal surface densities follow a specific profile from the start of the disc lifetime, and as such do not account for the formation of planetesimals.

To account for this disparity in size and mass between initial embryos and the formed planetesimals, pebble and planetesimal models assume that a period of runaway and subsequently oligarchic growth had occurred for the planetesimals that formed, allowing them to reach the initial embryo masses.
For the more massive planetesimals that form, they are able to grow significantly faster than their less massive counterparts, due to gravitational focusing as their collisional cross-sections are increased by a factor $\sqrt{1+v_{\rm esc}^2/v_{\rm rel}^2}$, where $v_{\rm esc}$ is the escape velocity from the more massive body and $v_{\rm rel}$ is the relative velocity of the planetesimals.
Once a small number of bodies become considerably more massive than the rest, such that their escape velocities dominate the average velocity in the disc, they undergo runaway growth quickly doubling their mass \citep{Wetherill89,Wetherill93,Kokubo96}.
Within the runaway growth regime, the doubling time is proportional to $M_{\rm p}^{-1/3}$, and continues until the more massive planetesimals begin to stir the velocity dispersion of the remaining planetesimals in the disc, such that $v_{\rm esc} \sim v_{\rm rel}$.
This regime is known as `oligarchic growth' \citep{Kokubo98}.
When large planetesimals become oligarchs, they no longer undergo runaway growth since the gravitational focusing effect is less efficient \citep{Kokubo98}.
In this regime the mass doubling time is proportional to $M_{\rm p}^{1/3}$ , slower than the runaway growth phase for the more massive bodies, and so the planetary embryos grow by accreting material in their feeding zones.
It is at this point of the planetesimals growth that most works consider these objects as their initial planetary embryos.

Recently, \citet{Lenz19} developed a method for planetesimal formation throughout a disc based on the evolution of dust and pebbles in the disc \citep{Birnstiel10}.
They assume that short-lived pressure bumps formed throughout the disc, akin to zonal flows \citep{Johansen12,Dittrich13,Bai2014}, and were able to trap dust and pebbles.
Once the dust-to-gas ratio exceeded unity and assuming that the pebble density was larger than the Roche density, then the trapped solids could undergo gravitational collapse forming planetesimals.
With this method accounting for planetesimal formation throughout the disc, it has significant advantages for planet formation over works that form planetesimals at specific locations of the disc \citep[e.g. the water iceline:][]{Drazkowska17,Liu19}.

In this work we examine the masses and sizes of \embs and planetesimals that form throughout an evolving protoplanetary disc.
We then examine the accretion trajectories of the formed \embs in the pebble accretion and planetesimal accretion scenarios, as well as a combined scenario where \embs can accrete pebbles and planetesimals concurrently.
The aim here is to examine what types of \embs form throughout the disc, and whether there are preferential regions where pebble or planetesimal accretion dominate their evolution.
This model could then be included in full planet formation simulations, giving more self-consistent initial conditions for the planetesimals and embryos in protoplanetary discs, that could then form the planets and planetary systems similar to those observed to this day.

This paper is organised as follows. In Sect. \ref{sec:model} we describe our 1D viscous disc model. Section \ref{sec:peb_to_pltml} details how planetesimals are formed from pebbles in our models.
In Sect. \ref{sec:embryos} we specify the properties of \embs that form from the largest planetesimals formed in Sect. \ref{sec:peb_to_pltml}.
We examine different accretion regimes for the newly formed \embs in Sect. \ref{sec:accretion}, and then we draw our conclusions in Sect. \ref{sec:conclusions}.

\section{Physical Model}
\label{sec:model}
\subsection{Gas disc}

We adopt a 1D viscous disc model where the equilibrium temperature is calculated by balancing irradiation heating from the central star, background heating from the residual molecular cloud, viscous heating and blackbody cooling.
The surface density, $\Sigma$, is evolving by solving the standard diffusion equation
\begin{equation}
\dfrac{d\Sigma}{dt}=\frac{1}{r}\dfrac{d}{dr}\left[3r^{1/2}\dfrac{d}{dr}\left(\nu\Sigma r^{1/2}\right)\right]-\dfrac{d\Sigma_{\rm pe}}{dt},
\end{equation}
where $\dfrac{d\Sigma_{\rm pe}}{dt}$ is the rate change in surface density due to photoevaporative winds, and $\nu$ is the disc viscosity \citep{Shak}
\begin{equation}
\nu=\alpha c_{\rm s}^2/\Omega,
\end{equation}
where $c_{\rm s}$ is the local isothermal sound speed, $\Omega = \sqrt{\frac{GM_*}{r^3}}$ is the Keplerian frequency and $\alpha$ is the viscosity parameter.
As the disc should be in thermal equilibrium, we use an iterative method to solve the following equation \citet{Dangelo12}
\begin{equation}
Q_{\rm irr} + Q_{\nu} + Q_{\rm cloud} - Q_{\rm cool} = 0,
\end{equation}
where $Q_{\rm irr}$ is the radiative heating rate due to the central star, $Q_{\nu}$ is the viscous heating rate per unit area of the disc, $Q_{\rm cloud}$ is the radiative heating due to the residual molecular cloud, and $Q_{\rm cool}$ is the radiative cooling rate.
For a Keplerian disc, the energy flux due to dissipation is given by \citet{Mihalas} as
\begin{equation}
Q_{\nu} = \frac{9}{4}\nu\Sigma\Omega^2.    
\end{equation}
The heating rate due to stellar irradiation is given by \citet{Menou}
\begin{equation}
Q_{\rm irr}=2\sigma T_{\rm irr}^4/\tau_{\rm eff}, 
\end{equation}
where
\begin{equation}
\tau_{\rm eff} = \frac{3\tau_{\rm R}}{8}+\frac{1}{2}+\frac{1}{4\tau_{\rm P}},
\end{equation}
where $\tau_{\rm R}$ and $\tau_{\rm P}$ are the optical depths due to the Rosseland and Planck mean opacities respectively (assumed to be equivalent in this work).
For the irradiation temperature we take
\begin{equation}
T_{\rm irr}=(T_{*}^4+T_{\rm acc}^4)(1-\epsilon_{\rm alb})\left(\frac{R_*}{r}\right)^2 W_{\rm G}.
\end{equation}
Here $\epsilon_{\rm alb}$ is the disc albedo (taken to be 0.5), $T_{\rm acc}$ is the contribution made to the irradiation temperature by the accretion of gas on to the star, $T_*$ and $R_*$ are the stars effective temperature and radius, and $W_{\rm G}$ is a geometrical factor that determines the flux of radiation that is intercepted by the disc surface.
This approximates to
\begin{equation}
W_{\rm G} = 0.4 \left(\frac{R_*}{r}\right) +\frac{2}{7}\frac{H}{r},
\end{equation}
as given by \citet{Dangelo12}.
The scale height of the disc is denoted by $H$ in the equation above and is equal to $c_{\rm s}/\Omega$.
For $Q_{\rm cloud}$ we have
\begin{equation}
Q_{\rm cloud} = 2\sigma T_{\rm cloud}^4/\tau_{\rm eff}
\end{equation}
where we take $T_{\rm cloud}$ as being equal to 10 K.
For the cooling of the disc we have
\begin{equation}
Q_{\rm cool} = 2\sigma T_{\rm mid}^4/\tau_{\rm eff}
\end{equation}
with $T_{\rm mid}$ being the disc midplane temperature.

\subsubsection{Opacities}
We take the opacity $\kappa$ to be equal to the Rosseland mean opacity, with the temperature and density dependencies calculated using the formulae in \citet{Bell97} for temperatures below 3730K, and by \citet{Bell94} above 3730K\footnote{For the purpose of these equations, where the opacity is dependant on the local gas density, a density of $10^{-9} \rm gcm^{-3}$ is used to calculate the temperature ranges where that opacity law is appropriate.}:
\begin{equation}
\kappa[cm^2/g] = \left\{ \begin{array}{ll}
10^{-4}T^{2.1} & T<132 \, {\rm K} \\
3T^{-0.01} & 132\le T<170 \, {\rm K} \\
T^{-1.1} & 170\le T<375 \, {\rm K} \\
5\times 10^{4}T^{-1.5} & 375\le T<390 \, {\rm K} \\
0.1T^{0.7} & 390\le T<580 \, {\rm K} \\
2\times 10^{15}T^{-5.2} & 580\le T<680 \, {\rm K} \\
0.02T^{0.8} & 680\le T<960 \, {\rm K} \\
2\times 10^{81}\rho T^{-24} & 960\le T<1570 \, {\rm K} \\
10^{-8}\rho^{2/3}T^{3} & 1570\le T<3730 \, {\rm K} \\
10^{-36}\rho^{1/3}T^{10} & 3730\le T<10000 \, {\rm K} \\
1.5\times 10^{20}\rho T^{-2.5} & 10000\le T<45000 \, {\rm K} \\
0.348 & T \ge 45000 \, {\rm K}
\end{array}\right.
\end{equation}
To account for changes in the disc metallicity, we multiply the opacity by the dust contribution to the metallicity relative to solar.

\subsection{Photoevaporation}

The absorption of UV radiation by the disc can heat the gas above the local escape velocity, and hence drive photoevaporative winds.
For extreme ultra-violet radiation (EUV), this creates a layer of ionised hydrogen with temperature $\sim$10,000~K \citep{Clarke2001}, whereas for far ultra-violet radiation (FUV), this creates a neutral layer of dissociated hydrogen with temperature of roughly 1000K \citep{Matsuyama03}.
We incorporate both EUV radiation from the central star (internal photoevaporation) and also FUV radiation from other nearby stars (external photoevaporation).
We do not include here the effects of X-ray induced internal photoevaporation \citep[e.g.][]{Owen12}, since they operate in the outer regions of the disc, similar to those where external photoevaporation operates, and with the interplay between internal and external photoevaporation being poorly understood, we choose to leave the inclusion of both effects to future work where we will examine the effects of different values within the observed parameter space for both internal and external photoevaporation rates.
The effects of FUV radiation from the central star are also neglected in this work, since it again operates in a similar location to FUV external photoevaporation, which we assume dominates the evolution of the disc in this region.
Whilst the internally originating FUV radiation is an important process, those models also strongly dependent on the local disc properties, e.g. the size of dust in the penetrated region of the disc \citep{Gorti15}, as well as complex photochemistry, including the photo- and chromo-spheres of the central stars \citep{Gorti09a,Gorti09}.
Including such complex models is beyond the scope of this paper, and will be subject to future work.

\subsubsection{Internal photoevaporation}
To account for the radiation from the central star we adopt the formula provided by \citet{Dullemond} to calculate the rate at which the surface density decreases due to this wind
\begin{equation}
\dfrac{d\Sigma_{\rm pe,int}}{dt} = 1.16\times10^{-11}G_{\rm fact}\sqrt{f_{41}}\left(\dfrac{1}{\rgeuv}\right)^{3/2}
\left(\dfrac{M_{\bigodot}}{\au^2 \, {\rm yr}}\right)
\end{equation}
where $G_{\rm fact}$ is a scaling factor defined as
\begin{equation}
G_{\rm fact} = \left\{ \begin{array}{ll}
\left(\dfrac{\rgeuv}{r}\right)^2 e^{\frac{1}{2}\left(1-\dfrac{\rgeuv}{r}\right)} 
& r\le \rgeuv, \\
\\
\left(\dfrac{\rgeuv}{r}\right)^{5/2} & r>\rgeuv.
\end{array} \right.
\end{equation}
Here, $\rgeuv$ is the characteristic radius beyond which gas becomes unbound from the system as a result of the EUV radiation, which is set to $10\au$ for Solar-mass stars, and $f_{41}$ is the rate at which extreme UV ionising photons are emitted by the central star in units of $10^{41}$ s$^{-1}$.

When the inner region of disc becomes optically thin, ionising photons can launch a wind off the inner edge of the disc, enhancing the photoevaporation rate.
The direct photoevaporation prescription that we adopt is taken from \citet{Alexander07} and \citet{Alexander09}, where the photoevaporative mass loss rate is given by 
\begin{equation}
\dfrac{d\Sigma_{\rm pe,int}}{dt}=2C_2\mu m_Hc_s\left(\dfrac{f_{41}}{4\pi\alpha_Bhr^3_{in}}\right)^{1/2}
\left(\dfrac{r}{r_{in}}\right)^{-2.42}.
\end{equation}
Here, $C_2=0.235$, $\alpha_{B}$ is the Case B recombination coefficient for atomic hydrogen at $10^4$K, having a value of $\alpha_B=2.6\times10^{-19}\text{m}^3\text{s}^{-1}$ \citep{Cox}, and $r_{\rm in}$ is the radial location of the inner disc edge.

\subsubsection{External photoevaporation}
In addition to EUV radiation from the central star photoevaporating the protoplanetary disc, there is also a contribution from the discs external environment.
This is typically considered to be the radiation that is emanating from newly formed stars, in particular young, hot, massive stars that release vast amount of high-energy radiation.
Here we include the effects of external photoevaporation due to far-ultraviolet (FUV) radiation emanating from massive stars in the vicinity of the discs \citep{Matsuyama03}.
This drives a wind outside of the gravitational radius where the sound speed in the heated layer is $T\sim$1000~K, denoted $\rgfuv$.
This leads to a reduction in the gas surface density as follows \citep{Matsuyama03}
\begin{equation}
\dfrac{d\Sigma_{\rm pe,ext}}{dt} =  \left\{ \begin{array}{ll}
0 & r\le \beta \rgfuv , \\
\\
\dfrac{\dot{M}_{\rm pe,ext}}{\pi(r_{\rm max}^2-\beta^2 \rgfuv^2)}& r>\beta \rgfuv.
\end{array} \right.
\end{equation}
where $\beta = 0.14$ \citep[similar to][]{AlexanderPascucci12} gives the effective gravitational radius that external photoevaporation operates above.
To ensure realistic disc lifetimes, we take the total rate $\dot{M}_{\rm pe,ext}$ to be equal to $10^{-7} M_{\bigodot}/yr$, consistent with the rates found in \citet{Haworth18} for discs around Solar mass stars in low $G_0$ environments.
Note that by also modifying the viscous alpha parameter, as well as the internal photoevaporation rate, realistic disc lifetimes can be obtained with weaker/stronger external photoevaporation rates.

\subsection{Active turbulent region}
Fully developed magnetohydrodynamic (MHD) turbulence is expected to arise in regions of the disc where the temperature exceeds 1000~K \citep{UmebayashiNakano1988, DeschTurner2015}. 
To account for the increased turbulent stress, we follow \citet{ColemanNelson16} in increasing the viscous $\alpha$ parameter when the temperature rises above 1000~K by using a smooth transition function
\begin{equation}
\alpha(r) = \left\{ \begin{array}{ll}
\alpha_{\rm visc} & r > r_{\rm act}, \\
\\
\alpha_{\rm visc} + \dfrac{\alpha_{\rm act}-\alpha_{\rm visc}}{2} \\
\times\left(\tanh\left(\dfrac{3(r_{\rm act}-r-5H(r))}{5H(r)}\right)+1\right) & r \le r_{\rm act}, \\
\end{array} \right.
\end{equation}
where $r_{\rm act}$ represents the outermost radius with temperature greater than 1000 K, $\alpha_{\rm visc}$ is the $\alpha$ in the disc where the temperature is below 1000 K, and $H(r)$ is the local disc scale height.
This smooth transition leads to a maximum $\alpha_{\rm act}$ in the hottest parts of the disc close to the central star.
The values we take for $\alpha_{\rm visc}$ and $\alpha_{\rm act}$ in our simulations are consistent with other works including an active region close to the star \citep{Flock19}, and can be found in Table. \ref{tab:discparameters}.

\subsection{Pebble model}
\label{sec:pebbles}
To account for the pebbles in the disc, we implement the pebble models of \citet{Lambrechts12,Lambrechts14} into our simulations, of which we briefly discuss below.
As a protoplanetary disc evolves, a pebble production front extends outwards from the centre of the system as small pebbles and dust grains fall towards the disc midplane, gradually growing in size.
Once the pebbles that form reach a sufficient size they begin to migrate inwards through the disc due to aerodynamic drag.
Following \citet{Lambrechts14}, the location of this pebble production front is defined as:
\begin{equation}
\label{eq:peb_front}
r_{\rm g}(t) = \left(\frac{3}{16}\right)^{1/3}(GM_*)^{1/3}(\epsilon_dZ_{0})^{2/3}t^{2/3}
\end{equation}
where $\epsilon_d = 0.05$ is a free parameter that depends on the growth efficiency of pebbles, whilst $Z_{0}$ is the solids-to-gas ratio.
Since this front moves outwards over time, this provides a constant mass flux of inwardly drifting pebbles equal to:
\begin{equation}
\label{eq:massflux}
\dot{M}_{\rm flux} = 2\pi r_{\rm g}\dfrac{dr_{\rm g}}{dt}Z_{\rm peb}(r_{\rm g})\Sigma_{\rm gas}(r_{\rm g})
\end{equation}
where $Z_{\rm peb}$ denotes the metallicity that is comprised solely of pebbles.
Combining the metallicity comprised solely of pebbles with that to which contributes to the remaining dust in the disc, gives the total metallicity of the system:
\begin{equation}
Z_{0} = Z_{\rm peb} + Z_{\rm dust}.
\end{equation}
Here we assume that 90$\%$ of the total metallicity is converted into pebbles, and that this ratio remains constant throughout the entire disc lifetime.
The remaining metallicity is locked up within small dust grains that contribute to the opacity of the disc when calculating its thermal structure, and again we assume this remains constant over time.
Assuming that the mass flux of pebbles originating from $r_{\rm g}$ is constant throughout the disc, we follow \citet{Lambrechts14} in defining the pebble surface density, $\Sigma_{\rm peb}$, as the following:

\begin{equation}
\label{eq:sigma_peb}
\Sigma_{\rm peb} = \dfrac{\dot{M}_{\rm flux}}{2\pi r v_r},
\end{equation}
where $v_r$ is the radial velocity of the pebbles equal to
\begin{equation}
\label{eq:vr}
v_r = 2\frac{\Stokes}{\Stokes^2+1}\eta v_{\rm K}-\frac{v_{\rm r,gas}}{1+\Stokes^2}
\end{equation}
\citep{Weidenschilling_77,Nakagawa86}, where $\Stokes$ is the Stokes number of the pebbles, $v_K$ is the local Keplerian velocity, $v_{\rm r,gas}$ is the gas radial velocity, and $\eta$ is the dimensionless measure of gas pressure support \citep{Nakagawa86},
\begin{equation}
\eta = -\frac{1}{2}h^2\dfrac{\partial ~{\rm ln} P}{\partial ~{\rm ln} r},
\end{equation}
where $h$ is the local disc aspect ratio.

As pebbles drift inwards, eventually they cross the water iceline, which we take as being where the local disc temperature is equal to 170 K.
Since pebbles are mostly comprised of ice and silicates, when they cross the iceline, the ices sublimate releasing trapped silicates, reducing the mass and size of the remaining silicate pebbles.
To account for the sublimation of ices, of which we assume comprise $50\%$ of the pebble mass, we multiply the mass flux of pebbles drifting through the disc at radial locations interior to the iceline by a factor of 0.5 \citep{Lambrechts14b}.

\subsection{Example of the disc evolution}
\label{sec:disc_example}

\begin{table}
\centering
\begin{tabular}{lc}
\hline
Parameter & Value\\
\hline
Disc inner boundary & 0.04 $\au$\\
Disc outer boundary & 200 $\au$\\
Disc mass & 0.1 $\rm {M_*}$\\
Initial $\Sigma_{\rm g}$(1 $\au$) & $840~ {\rm gcm}^{-2}$\\
Initial surface density exponent & -1\\
Metallicity & 0.01\\
$\alpha_{\rm visc}$ & $1 \times 10^{-3}$\\
$\alpha_{\rm act}$ & $5 \times 10^{-3}$\\
$M_*$ & $1 \rm M_{\bigodot}$\\
$R_*$ & $2 \rm R_{\bigodot}$\\
$T_*$ & 4280 K\\
f41 & 10\\
$\dot{M}_{\rm pe, ext}$ & $10^{-7} {\rm M_{\bigodot}}/yr$\\
\hline
\end{tabular}
\caption{Disc and stellar model parameters}
\label{tab:discparameters}
\end{table}

\begin{figure*}
\centering
\includegraphics[scale=0.4]{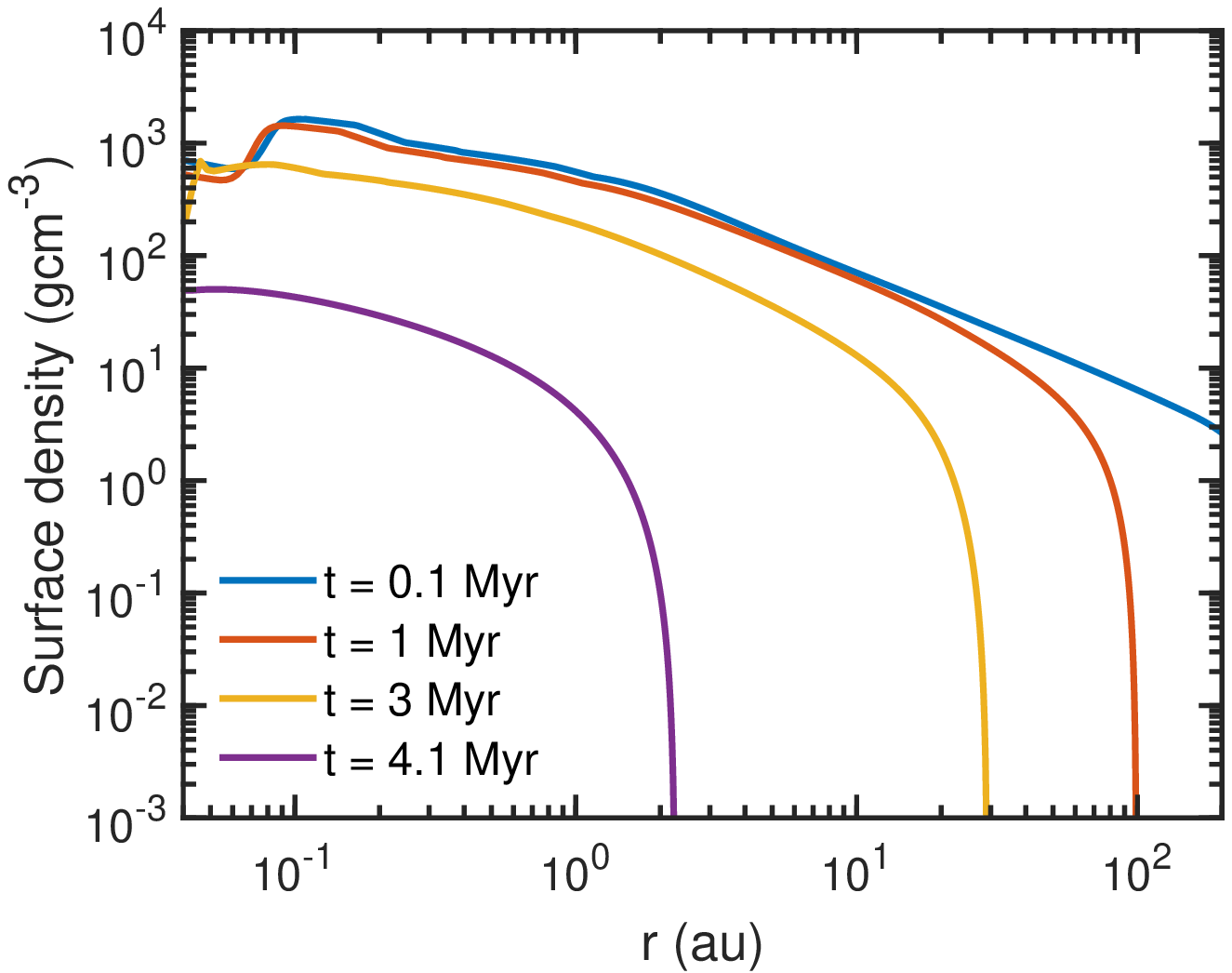}
\hspace{0.5cm}
\includegraphics[scale=0.4]{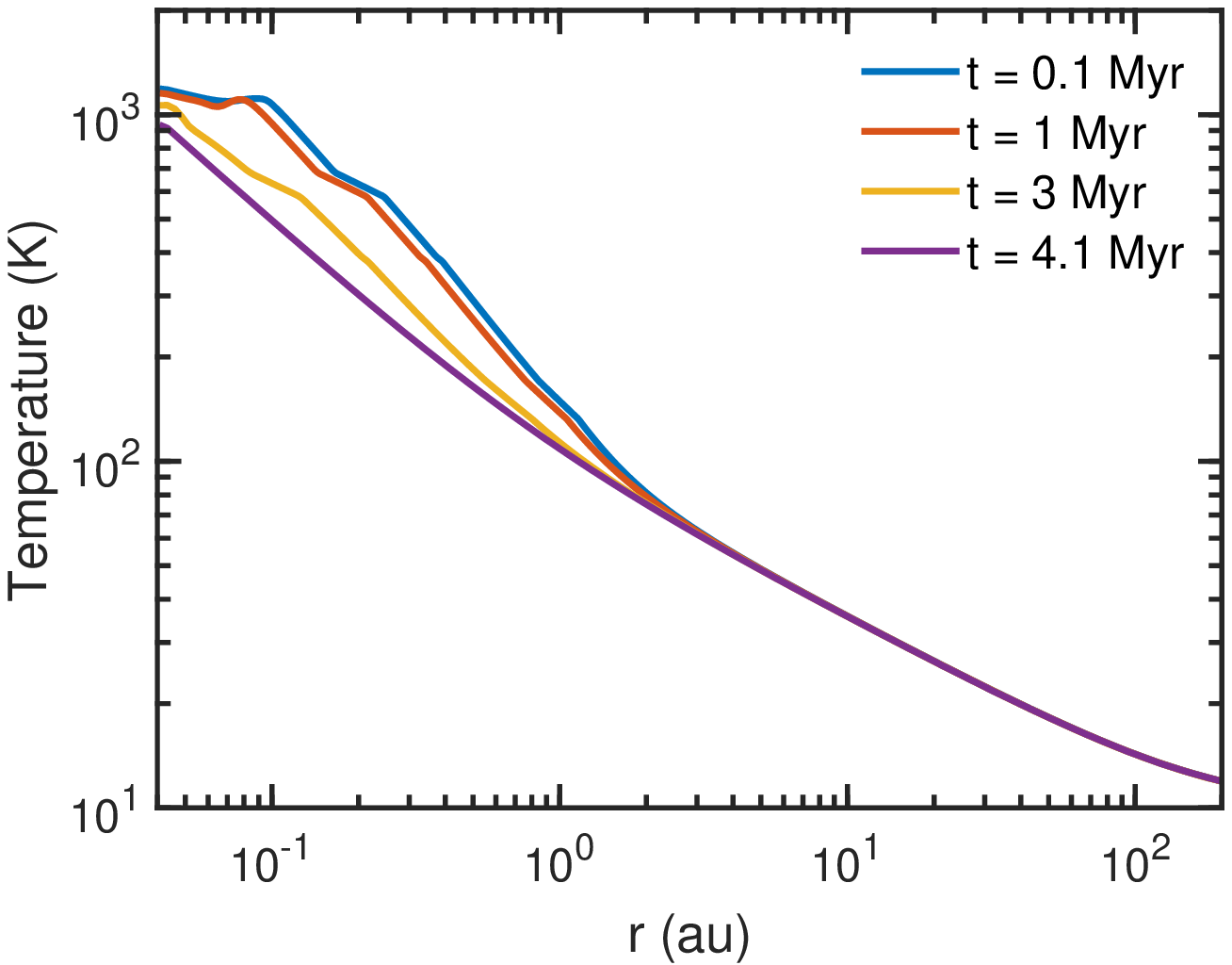}
\hspace{0.5cm}
\includegraphics[scale=0.4]{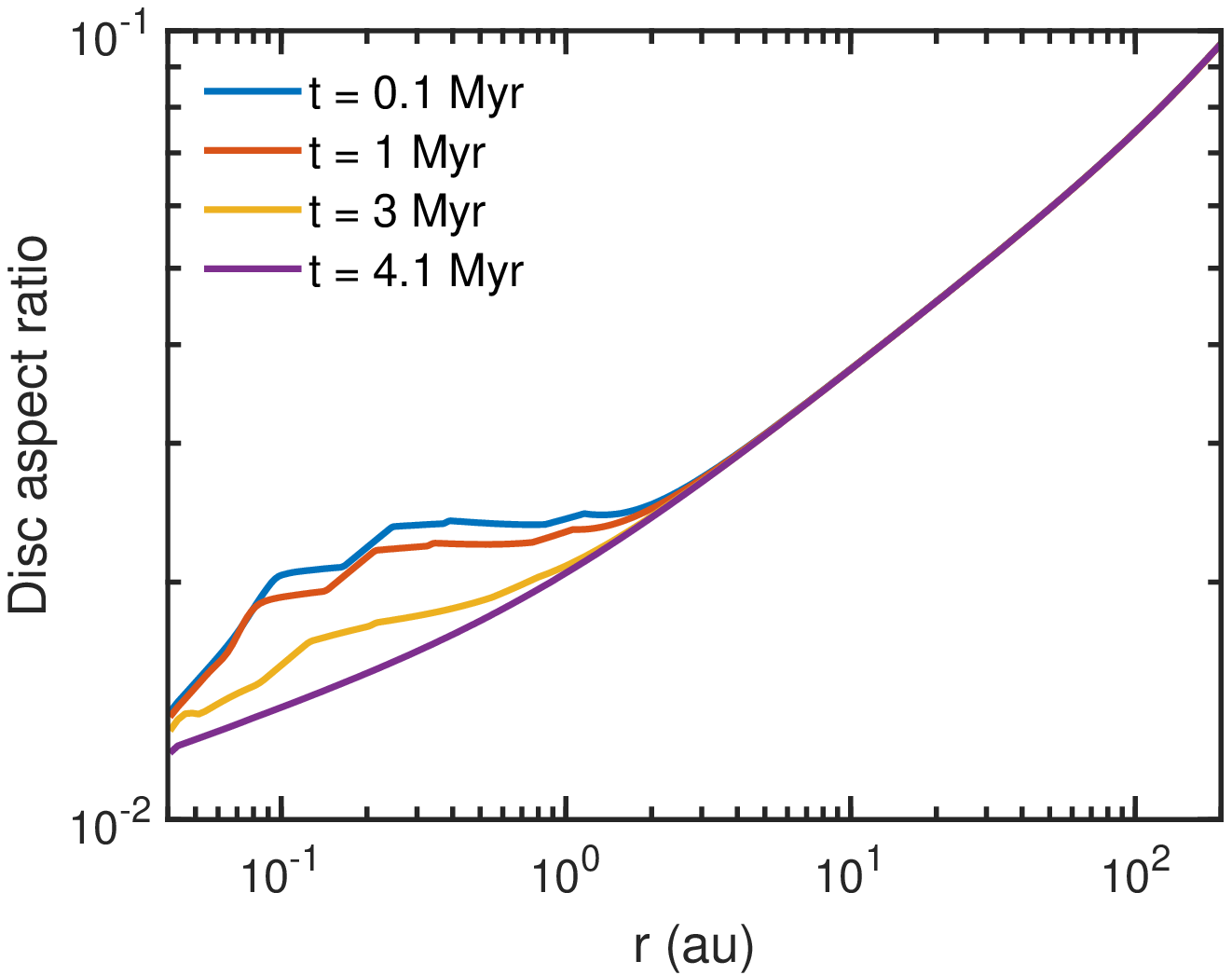}
\caption{Gas surface densities, temperatures and aspect ratios after 0.1, 1, 3 and 5 Myr (top-bottom lines) of the fiducial protoplanetary disc.}
\label{fig:multiplot}
\end{figure*}

Where the sections above outlined the ingredients of the physical model, we now describe the evolution of the protoplanetary disc.
Table \ref{tab:discparameters} gives the disc parameters used in the model.
The disc had an initial mass of $10\%$ that of the star, and with the parameters used had a lifetime of 4.1 Myr, compatible with observed disc lifetimes of between 3 and 10 Myr \citep{Mamajek09,Ribas14}.

Figure \ref{fig:multiplot} shows the evolution of the disc, with the disc surface density profiles in the left-hand panel, temperature profiles in the middle panel, and $H/r$ profiles in the right-hand panel.
As time progresses the inner gas disc viscously accretes inwards and eventually on to the central star, gradually reducing the surface density over time as is seen by different temporal profiles in fig \ref{fig:multiplot}.
The outer disc viscously spreads outwards and is evaporated by high-energy radiation emanating from external sources.
This allows for the outer edge of the disc to being able to be dynamically controlled by the local viscosity and the external photoevaporation rate, as can be seen by the yellow line in the left-hand panel of fig \ref{fig:multiplot}.
Internal photoevaporation from the central star also reduces the surface density over the majority of the disc as it evolves.
The dip in the surface density profiles close to the inner edge of the disc ($r\le 0.1\au$) are a result of the active turbulent region, where T$>1000$K causes an increase in the viscosity.
Over time, this region moves in towards the central star as the reduction in surface density reduces the viscous heating rate and opacity, as can be seen with the evolution of the blue to the yellow lines.
The turbulent region disappears when the disc temperature no longer exceeds 1000 K anywhere in the disc, as is shown by the yellow and purple lines in the middle plot of fig. \ref{fig:multiplot}.
At the end of the disc lifetime, the outer edge of the disc has receded to less than 2 au through the viscous accretion of material towards the inner disc, and through outward viscous spreading fuelling internal and external photoevaporative winds.
For these disc parameters, there is no inner hole that appears due to internal photoevaporation in contrast to other works \citep{Alexander07,Gorti15,ColemanNelson16}.
This is due to the external photoevaporative exhausting the supply of gas in the outer disc that for discs without external photoevaporation or with a significantly weaker external rate, would continue to supply sufficient gas for internal photoevaporation to eventually be able to open a hole in the disc similar to previous works.

\section{Pebbles To Planetesimals}
\label{sec:peb_to_pltml}
\subsection{Conditions required for planetesimal formation}
Now that sect. \ref{sec:model} has outlined the gas and pebble model in the disc, with sect. \ref{sec:disc_example} showing the gas disc's evolution, we now concentrate on the conversion of pebbles into planetesimals.
It is typically assumed that planetesimals form through the gravitational collapse of smaller dust and pebbles, when the local particle density exceeds the Roche density, which can occur when the dust-to-gas ratio exceeds unity \citep{Johansen07,JohansenYoudin2009}.
One way of achieving a dust-to-gas ratio of unity is to concentrate dust and pebbles at pressure bumps in the disc.
These pressure bumps have been observed to form in numerous local \citep{Johansen09,Simon12,Dittrich13} and global \citep{SteinackerPapaloizou2002,PapaloizouNelson2003,FromangNelson2006} magnetohydrodynamic (MHD) simulations, and more recently in simulations including non-ideal MHD effects \citep{Bai2014,ZhuStoneBai2014,BethuneLesur2016}.
They typically arise from localised magnetic flux concentration and the associated enhancement of magnetic stresses.

The main problem in forming the planetesimals is consistently attaining locations in the disc where these pressure bumps form.
One such location is the water iceline, where the change in gas opacities, and the local composition of the gas disc can create a pressure bump \citep{Drazkowska17}.
However forming planetesimals at the water iceline does not explain the formation of planetesimals at larger orbital radii, e.g. the Kuiper belt, nor does it facilitate in the formation of planets further out in the disc, e.g. Jupiter or Saturn.
If there were significant outward migration in the disc, then the problem of only forming planetesimals at the water iceline could be overcome since the planetary cores could undergo significant outward migration \citep{Paardekooper2014,McNally18}.

Whilst it may appear difficult to form long-lived pressure bumps throughout the entire protoplanetary disc, it would be reasonable to assume that short-lived local pressure bumps could form stochastically throughout the disc, enabling local dust-to-gas ratios to exceed unity for short times only.
This approach was recently examined by \citet{Lenz19} where they assumed that pressure bumps formed throughout the protoplanetary disc, creating traps for solids that would then collapse and form planetesimals.
We follow the approach of \citet{Lenz19} in forming our planetesimals, and we outline this approach below.

Whilst we assume that the pebble traps are forming throughout the disc, they do possess a significant lifetime before they dissipate.
We take this lifetime to be equal to 100 local orbital periods, with their formation/dissipation assumed to occur almost instantaneously.
Given that it takes significant time for all of the pebbles at a location to fall to the midplane and begin drifting towards the centre of the system, we only allow the pebble traps to begin converting the trapped pebbles into planetesimals 100 local orbital periods after the pebble growth front has reached a location $r$ (eq. \ref{eq:peb_front}).
We then only assume that a specific percentage of pebbles are trapped in the pebble traps, $\epsilon$, with the remainder drifting past.
This could be the case if the pebble traps are not fully azimuthally encompassing, i.e. vortices, and as such there are areas of the disc azimuthally that do not hinder the inward drift of pebbles.
We therefore define $\Sigma_{\rm trap}$ as:
\begin{equation}
\label{eq:sig_trap}
    \Sigma_{\rm trap} = \dfrac{\epsilon M_{\rm flux} \tau_{100}}{2\pi r l}
\end{equation}
where $\tau_{100}$ is equal to 100 local orbital periods, and $l$ is the length scale over which planetesimal formation occurs within the pebble trap, which we take to be equal to $0.01 H_{\rm gas}$ \citep[][see their eq. (3.40)]{Schreiber18}.

Just because the pebble traps are able to trap pebbles, this doesn't necessarily mean that the planetesimals are able to form there.
For planetesimals to form through for example gravitational collapse following the streaming instability, we assume that the local dust-to-gas ratio at the disc midplane has to be equal to or exceed unity, i.e. the dust density has to be greater than or equal to the gas density at the disc midplane \citep{Youdin05}.
When the dust-to-gas ratio exceeds unity, the growth rates of the fastest growing modes significantly increase, allowing for particles to concentrate on short time-scales, which could then undergo gravitational collapse.
Following \citet{Youdin07}, we define $H_{\rm peb}$ as
\begin{equation}
    H_{\rm peb} = H_{\rm gas}\sqrt{\dfrac{\alpha}{\Stokes}}
\end{equation}
where $\Stokes$ is the Stokes number and assumed to be equal to,
\begin{equation}
\Stokes = {\rm min}~(\Stokes_{\rm drift}, \Stokes_{\rm frag}),
\end{equation}
where $\Stokes_{\rm drift}$ is the drift-limited Stokes number that is obtained through an equilibrium between the drift and growth of pebbles to fit constraints of observations of pebbles in protoplanetary discs and from advanced coagulation models \citep{Birnstiel12},
\begin{equation}
    \Stokes_{\rm drift} = \frac{\sqrt{3}}{8}\frac{\epsilon_{\rm d}}{\eta}\frac{\Sigma_{\rm peb}}{\Sigma_{\rm gas}}.
\end{equation}
As well as the drift-limited Stokes number, we also include the fragmentation-limited Stokes number, $\Stokes_{\rm frag})$ which is equal to
\begin{equation}
    \Stokes_{\rm frag} = \frac{v_{\rm frag}^2}{3 \alpha c_{\rm s}^2},
\end{equation}
where $v_{\rm frag}$ is the impact velocity required for fragmentation, which we model as the smoothed function,
\begin{equation}
    \frac{v_{\rm frag}}{\rm 1m/s} = 10^{0.5+0.5 \tanh((r-r\rm snow)/5H)}.
\end{equation}

Then by equating the pebble density to the gas density, we derive the pebble surface density required for the streaming instability to occur, $\Sigma_{\rm SI}$,
\begin{equation}
\label{eq:sig_SI}
\Sigma_{\rm SI} = \Sigma_{\rm gas}\sqrt{\dfrac{\alpha}{\Stokes}}.
\end{equation}
For gravitational collapse to be able to occur, $\Sigma_{\rm trap} \ge \Sigma_{\rm SI}$.

A further condition for planetesimals to form through gravitational collapse is that the local midplane density of pebbles has to be larger than the Roche density, $\rho_{\rm p} > \rho_{\rm Roche}$, where
\begin{equation}
    \rho_{\rm Roche} = \frac{9}{4\pi}\frac{M_*}{r^3}
\end{equation}
When this condition is met, the self-gravity of the pebbles is strong enough to overcome the Keplerian shear which leads to the gravitational collapse of the pebbles \citep{GoldreichWard73,Johansen14}.
Therefore with this criterion, it means that even when the pebble density is greater than the gas density and stable filaments are effectively formed, those filaments will not produce any planetesimals unless $\rho_{\rm p}>\rho_{\rm Roche}$.
Assuming that the pebble density inside the filaments is proportional to the gas density, we have
\begin{equation}
    \rho_{\rm p,SI} = \epsilon_{\rm R}\rho_{\rm gas}
\end{equation}
with $\rho_{\rm p,SI}$ being the pebble density within a filament produced by the streaming instability,  $\rho_{\rm gas}$ being the gas midplane density, and $\epsilon_{\rm R}$ being an enhancement factor.
As the streaming instability concentrates the pebbles into denser and denser filaments, the value of $\epsilon_{\rm R}$ will increase such that $\epsilon_{\rm R} \gg 1$ inside the filaments.

Recent studies have found varying values for $\epsilon_{\rm R}$ with simulations using different Stokes numbers, as well as local disc metallicities $Z = \Sigma_{\rm p}/\Sigma_{\rm g}$.
For particles of $\Stokes=0.3$, and $Z=0.02$, \citet{Johansen15} found a a local pebble enhancement of $10^4$ times the local gas density.
Using smaller particles with $\Stokes = 10^{-2}$ and $\Stokes = 10^{-3}$ with $Z=0.02$--$0.04$, \citet{Yang17} observed concentrations between 10 and $10^3$.
More recently \citet{Carrera21} found concentrations of between $10^2$ and $10^4$ for Stokes numbers ranging between 0.07 and 0.3 in discs with Z=0.01.
Given that it is still unclear of how $\epsilon_{\rm R}$ depends on the Stokes number and the local disc metallicity, as well as possibly other simulation parameters, we assume that $\epsilon_{\rm R}$ to be equal to $10^4$ as within our simulations, the Stokes number typically ranges between 0.05 and 0.2.\footnote{Note that taking $\epsilon_{\rm R}=10^3$ yields negligible differences in our results, since the Roche density criterion only becomes significant in the inner $\sim$few $\au$ of the disc.}

Therefore for planetesimals to be formed, we require $\Sigma_{\rm trap} \ge \Sigma_{\rm SI}$, that is when the pebble midplane density will be greater than the gas midplane density, and $\epsilon_{\rm R}\rho_{\rm gas} > \rho_{\rm Roche}$, that is the pebble midplane density within a filament is greater than the Roche density.
Once these conditions are met, we then remove the mass of pebbles that are converted to planetesimals and apply it to the formed planetesimal surface density $\Sigma_{\rm pltml}$
\begin{equation}
\label{eq:sig_pltml}
    \dot{\Sigma}_{\rm pltml}(r) = \Sigma_{\rm trap}\dfrac{\Delta t}{\tau_{100}}
\end{equation}
where $\Delta t$ is the time step.
We then modify eq. \ref{eq:massflux} to remove the appropriate mass from the mass flux heading further downstream in the system
\begin{equation}
    \dot{M}_{\rm flux} = \dot{M}_{\rm flux,0} - \dot{M}_{\rm pltml}
\end{equation}
before this mass flux is then used to calculate the local pebble surface density (see eq. \ref{eq:sigma_peb}), with $\dot{M}_{\rm flux,0}$ being the initial mass flux of pebbles without the the inclusion of planetesimal formation.

\begin{figure}
\centering
\includegraphics[scale=0.5]{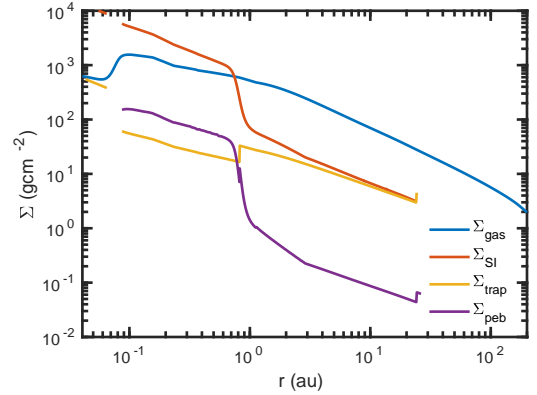}
\caption{Surface densities as a function of orbital distance after 0.1 Myr for: gas (blue, top line), required solids surface density for the streaming instability (eq. \ref{eq:sig_SI}, red line), amount of solids trapped in a short-lived pressure bump (eq. \ref{eq:sig_trap}, yellow line), and pebble (purple, bottom line).}
\label{fig:sigmas}
\end{figure}

\begin{figure}
\centering
\includegraphics[scale=0.5]{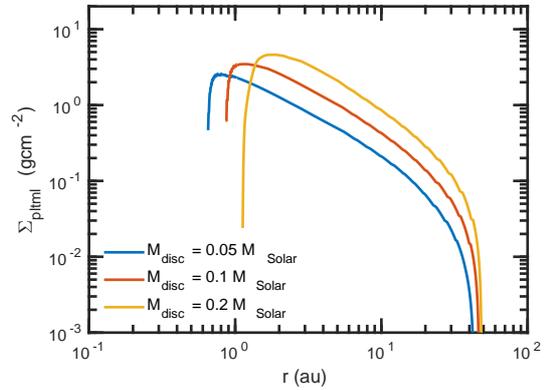}
\caption{The final planetesimal surface densities for protoplanetary discs of mass: $0.05 \msun$ (blue line), $0.1 \msun$ (red line), and $0.2 \msun$ (yellow line).}
\label{fig:sigma_pltml}
\end{figure}

The profiles for the gas surface density $\Sigma_{\rm gas}$, the surface density required for gravitational collapse to occur $\Sigma_{\rm SI}$, the surface density of trapped pebbles $\Sigma_{\rm trap}$ and the resulting pebble surface density $\Sigma_{\rm peb}$ are shown in fig. \ref{fig:sigmas} for our fiducial model.
These profiles are shown at a time of 0.1 Myr into the simulation.
The pebble growth front, $r_{\rm g}$ is located at approximately $25 \au$ at this time, and can be seen to the far right of the pebble surface density profile.
The sharp drop in the pebble surface density just interior to that location is where the pebbles are being converted to planetesimals on short time-scales at one of the pebble traps.
This can be seen where pebble trap surface density (yellow line) is larger than the surface density required for gravitational collapse to occur (red line).
Moving inwards in the disc, the pebble trap surface density now sharply drops due to the reduction in the amount of pebbles that are available to be trapped.
This continues all the way to the inner regions of the disc, with the trapped pebbles unable to meet the conditions for gravitational collapse.
If at other times of the simulation, the pebble trap surface density in the inner disc did exceed that required for gravitational collapse, then a number of planetesimals would indeed be formed at that location, even though the pebble growth front, and the main planetesimal formation location may be much further away in the outer regions of the disc.

Over time, the yellow, red, and purple profiles seen in fig. \ref{fig:sigmas} move outwards as the pebble growth front progress further out into the disc.
With the reduction in the gas surface density over time, as seen in fig. \ref{fig:multiplot}, the profiles also decrease as time progresses due to the reduction in pebbles formed over time.
Eventually the lack of pebbles formed has an impact on the pebble trap surface density, where at no locations is $\Sigma_{\rm trap}$ greater than $\Sigma_{\rm SI}$, and as such no more planetesimals are assembled.
For our fiducial disc, this occurs after $\sim 0.3$ Myr with the pebble growth front being located exterior to $50 \au$.
After this, assuming no further gravitational collapses occur, and no planetesimals are removed from the disc through accretion, drift or ejection, then the planetesimal surface densities will remain constant.

Figure \ref{fig:sigma_pltml} shows the planetesimal surface densities for our fiducial disc (red line) as well as for discs with masses equal to 0.05$\msun$ (blue line) and 0.2$\msun$ (yellow line).
The dips located between 0.5--2$\au$ in the disc profiles show the location of the water iceline in the respective models, situated at larger orbital radii in the more massive discs due to their increases in viscous heating.
Just exterior to the iceline, the pebble surface density begins to increase significantly.
This is due to the pebbles substantially reducing in size as the collision velocities between pebbles exceeds the fragmentation velocities, resulting in the Stokes number being fragmentation limited.
Since the pebbles are smaller in size and subsequently Stokes number, they are more greatly coupled to the gas, and as such, drift inward through the disc more slowly, resulting as per eq. \ref{eq:sigma_peb} in a larger pebble surface density.
The reduction in pebble size is further enhanced around the iceline as the pebbles undergo sublimation, losing their water components, leaving only the silicate remnants.
With the Stokes number of the pebbles being much reduced, the pebble scale height is significantly larger.
This results in the surface density required for gravitational collapse to be much larger then the surface density for pebble trapping, ultimately leading to no planetesimals being formed within the iceline in these models.
This can be seen on the left-hand side of fig \ref{fig:sigma_pltml} where the planetesimal surface density interior to the iceline is zero.
The lack of planetesimal formation raises questions about whether planets such as Mercury can form {\it{in situ}} or whether they need to form at or exterior to the iceline and migrate closer to their central stars \citep{Johansen21}.

\begin{figure}
\centering
\includegraphics[scale=0.5]{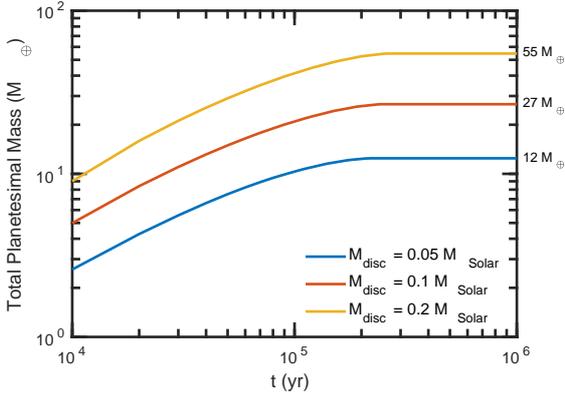}
\caption{The total planetesimal mass as a function of time for protoplanetary discs of mass: $0.05 \msun$ (blue line), $0.1 \msun$ (red line), and $0.2 \msun$ (yellow line). The final masses are indicated at the right of the figure.}
\label{fig:mpltml_time}
\end{figure}

In terms of the total mass in planetesimals formed, our fiducial disc model contains $27 \me$ at the end of the disc lifetime.
Figure \ref{fig:mpltml_time} shows the growth in the total planetesimal mass throughout the first 1 Myr for our fiducial simulation (red line) and the simulations with initial disc masses of $0.05 \times \msun$ (blue line) and $0.2 \times \msun$ (yellow line).
As can be seen by the final planetesimal masses on the right-hand side of fig. \ref{fig:mpltml_time}, the doubling of the initial disc mass, results in an approximate doubling of the total mass in planetesimals formed.
This is unsurprising since the mass flux of pebbles formed at pebble production front, as well as the pebble surface density required for the streaming instability, are proportional to the local gas surface density (eq. \ref{eq:massflux}).
This results in the production of planetesimals occurring radially at similar rates, but at increasing magnitudes depending on the initial disc mass.
After 0.3 Myr, the formation of planetesimals ends due to the mass of pebbles being trapped being insufficient to undergo gravitational collapse, i.e. $\Sigma_{\rm trap} < \Sigma_{\rm SI}$.
For accreting \embs, this can increase their accretion rate, since the pebble mass flux passing them would be slightly increased compared to a slightly earlier time in their evolution.

In comparison to other works, the final mass comprised within planetesimals in our fiducial model is comparable to that found in \citet{Voelkel20} who use a similar method based on the models of \citet{Lenz19}.
In comparing to \citet{Lenz19} the total planetesimal mass in our equivalent disc (of mass $0.05\msun$) is significantly lower than that found in \citet{Lenz19} ($12 \me$ versus $\sim100\me$).
This difference is expected due to the different initial disc profiles and subsequent evolution, as well as differing criteria for when planetesimals form.
They assume that a planetesimal forms when the mass of pebbles that become trapped is equal to the mass of a planetesimal of 100km in size, irrespective of whether the local disc conditions favour the gravitational collapse of the trapped pebbles to form such an object.

Recently \cite{Eriksson20} examined the formation of planetesimals in the presence of embedded planets in massive protoplanetary discs, finding that planetesimal formation was efficient at the pressure bumps formed exterior to their embedded planet's orbits.
In total their simulations typically formed $>200 \me$ of planetesimals, a value much larger than those formed in this work, however their initial disc mass was considerably larger than that examined in this work (an order magnitude than our fiducial model), and as such by applying the simple mass scaling seen in fig. \ref{fig:mpltml_time} and using a similar mass to that used in \citet{Eriksson20}, we would obtain comparable masses of formed planetesimals.
When comparing to the most complete model of \citet{Carrera17} who examine planetesimal formation at the late stages of a photoevaporating protoplanetary disc, the total planetesimal masses formed in this work are a factor 2-7 times smaller than found there.
However when looking at the inner 100 $\au$, the location where planetesimals generally form in this work, the masses become comparable, on order of tens of Earth masses.
In \citet{Carrera17}, the majority of their planetesimals form in the outer regions of the disc beyond 100 $\au$ where the local disc metallicity is more favourable in facilitating planetesimal formation, and as such most of their planetesimals form at large distances.
This does not occur to such an extent in the models in this work, since we use a more simple pebble model and assume that the pebbles quickly drift inwards once their drift time-scales become comparable to their growth time-scales.
In addition to the differences in planetesimal formation locations, the planetesimals formed in the inner regions in \citet{Carrera17} only form late in the disc lifetime, after 2.5 Myr.
Given this lateness, it raises questions as to whether there is sufficient time for planets to form from such a planetesimal reservoir before the end of the disc lifetime.

\begin{figure}
\centering
\includegraphics[scale=0.5]{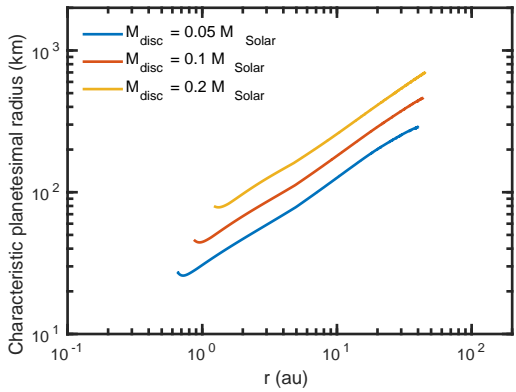}
\caption{The characteristic planetesimal radius (eq. \ref{eq:r_0}) as a function orbital distance for protoplanetary discs of mass: $0.05 \msun$ (blue line), $0.1 \msun$ (red line), and $0.2 \msun$ (yellow line).}
\label{fig:characteristic_radius}
\end{figure}

\subsection{Planetesimal sizes and masses}
The planetesimals that form through gravitational collapse have a specific size.
Typically it was assumed that the initial planetesimal size was around 100 km.
This was based on planetesimal formation models \citep{Youdin05,Johansen07,JohansenYoudin2009,Bai10} as well as on observations of the Solar System \citep{Morbidelli09,Delbo17}.
This is also the assumed planetesimal size in other planetesimal formation models \citep{Lenz19,Lenz20,Voelkel20}.
However more recent studies of gravitational collapse in pressure bumps in protoplanetary discs have found that the characteristic planetesimal size on the order of a few hundred kilometres \citep{Johansen12,Johansen15,Simon16,Schafer17,Simon16,Simon17,Abod19,Li19}.
It has also been found more recently to be dependent on the local disc properties \citep{Simon17,Abod19,Li19}, which for constant densities, results in an increase in planetesimal radius as a function of orbital distance.

Numerous works have found that the initial mass function of streaming-derived planetesimals can be roughly fitted by a power-law plus an exponential decay \citep{Johansen15,Schafer17,Abod19}.
Recently \citet{Abod19} showed that the initial mass function depends only weakly on the aerodynamic properties of the disc and participating solids, when they cincluded the effects of particle self-gravity within their streaming instability simulations.
Following \citet{Abod19}, the initial mass of function of planetesimals is,
\begin{equation}
\label{eq:N_mpltml}
N(>M_{\rm pltml}) = C_1 M_{\rm pltml}^{1-p} \exp[-M_{\rm pltml}/M_0]
\end{equation}
where $p \simeq 1.3$, and $C_1$ is the normalisation constant set by the integrated probability equalling unity \citep{Meerschaert12},
\begin{equation}
    C_1 = M_{\rm pltml, min}^{p-1}\exp[{M_{\rm pltml, min}/M_0}], 
\end{equation}
where $M_{\rm pltml, min}$ is the minimum planetesimal mass formed by the streaming instability (taken in this work as being equal to $0.01\times M_0$).
The characteristic mass $M_0$ denotes the mass where the initial planetesimal mass function begins to steepen with the exponential part of eq. \ref{eq:N_mpltml} beginning to dominate.
Therefore $M_0$ can be treated as a proxy for the maximum planetesimal size and given that the majority of the mass in planetesimals that form through the streaming instability is tied up in the most massive planetesimals, we assume that $M_0$ is the average mass and therefore size of the planetesimals that form in our simulations.
Given that the tail of the initial mass function is exponential, it is possible to form planetesimals more massive than $M_0$.
In their simulations, \citet{Abod19} find that the most massive planetesimals formed are of the order of the gravitational mass, that is the maximum mass where self-gravity forces are stronger than the tidal shear forces emanating from the particle clumps interactions with the local disc, typically an order of magnitude larger than the characteristic mass.
As well as the work of \citet{Abod19}, other works have also found similar results regarding the initial planetesimal mass function and expressions for calculating the characteristic mass $M_0$ \citep{Johansen15,Schafer17,Simon16,Li19}.
More recently, \citet{Liu20} extrapolated on the simulations from those works, and derived an expression for the characteristic planetesimal mass,
\begin{equation}
\label{eq:MG_Liu}
M_0 = 5\times 10^{-5}\left(\frac{Z}{0.02}\right)^{1/2}\left(\pi\gamma\right)^{3/2}\left(\frac{h}{0.05}\right)^3\left(\frac{M_*}{1\msun}\right)\me,
\end{equation}
where $Z=\Sigma_{\rm peb}/\Sigma_{\rm gas}$ is the local disc metallicity and $\gamma=4\pi G\rho_{\rm gas}/\Omega$ is a self-gravity parameter.
Here we take $\Sigma_{\rm peb}$ to be equal to $\Sigma_{\rm SI}$, that is the surface density of pebbles required for the pebble cloud to undergo gravitational collapse.
We use $\Sigma_{\rm SI}$ instead of $\Sigma_{\rm trap}$ as we assume that once the requisite mass in pebbles has become trapped, i.e. $\Sigma_{\rm SI}$, then the pebble cloud undergoes gravitational collapse.
This is valid, so long as the pebble midplane density is larger than the Roche density, which is the case when $\Sigma_{\rm SI}>\Sigma_{\rm trap}$, except for the innermost region of the disc, well inside in the iceline.
Assuming a density $\rho_{\rm pltml}$ of the planetesimal, we then convert the characteristic mass, $M_0$ into a characteristic radius for the planetesimals
\begin{equation}
\label{eq:r_0}
    r_{\rm 0} = \left(\frac{3M_0}{4\pi\rho_{\rm pltml}}\right)^{1/3}.
\end{equation}

Figure \ref{fig:characteristic_radius} shows the characteristic planetesimal radius, $M_0$ as a function of orbital distance for the discs of different initial masses.
For the planetesimal density, we assume that the density is equal to $5~\rm gcm^{-3}$ interior to the water iceline, and $2~\rm gcm^{-3}$ exterior to the water iceline (assumed here when the local disc temperature is equal to 170 K).
When looking at the differences radially, it is clear to see that the planetesimal radius increases with orbital distance.
This is unsurprising, since when expanding eq. \ref{eq:r_0}, we find
\begin{equation}
\label{eq:r_0_propto}
r_0 \propto \Sigma_{\rm peb}^{1/6} \Sigma_{\rm gas}^{1/3} h^{1/2} r^{1/4},
\end{equation}
resulting in larger planetesimals at larger orbital radii for our disc models.

In our fiducial model (red line), planetesimal radii around the water iceline are of order 40km, whilst at 4 $\au$, they are around 100 km, 180km at 10 $\au$ and 450 km at 40 $\au$, consistent with the larger asteroid sizes in the Main Asteroid Belt as well as the Kuiper Belt.
These variations in sizes will have important consequences for planetesimal accretion scenarios, since they will affect the manoeuvrability of the planetesimals in the disc \citep{Adachi,Weidenschilling_77}, as well as their accretion rates onto \embs \citep{Inaba}.
It is also worth noting that these planetesimal sizes are much larger than what is currently used in planetesimal accretion scenarios \citep[e.g.][]{Mords09,ColemanNelson14,Mordasini15,ColemanNelson16,ColemanNelson16b}, and so significant collisional evolution would need to occur for the planetesimals to be ground down to the requisite sizes for global planetesimal accretion scenarios to be efficient.

When comparing the characteristic planetesimal radius from different disc masses, there is a clear increase in size for larger disc masses.
This again is not unexpected since eq. \ref{eq:r_0_propto} shows that as the pebble and gas surface densities increase, so will the planetesimal radii.

\section{Proto-Embryo Formation}
\label{sec:embryos}

Now that sects. \ref{sec:pebbles} and \ref{sec:peb_to_pltml} have described how the pebbles and planetesimals are formed throughout the protoplanetary disc, we now move onto the formation of \embs.
When the planetesimals form through gravitational collapse, numerous works found relations for the planetesimal initial mass functions \citep{Johansen15,Schafer17,Abod19}.
Whilst the number of massive planetesimals drops exponentially at masses greater than the characteristic mass, should there be enough mass being converted from pebbles to planetesimals, then a number of more massive planetesimals will be able to form.
It is these more massive planetesimals that we assume to be the \embs that will eventually grow into the planets that are observed in planetary systems.
In our simulations, we assume that for each planetesimal formation event, the largest planetesimal that forms is a \emb.
Using eq. \ref{eq:N_mpltml} we calculate the largest single body that forms in each event, (i.e. $N(>M_{\rm emb})=1$).
The mass of this \emb is then removed from the local mass of planetesimals over the lifetime of the traps, resulting in eq. \ref{eq:sig_pltml} becoming
\begin{equation}
\label{eq:sig_pltml2}
    \dot{\Sigma}_{\rm pltml}(r) = \left(\Sigma_{\rm trap}-\frac{M_{\rm emb}}{2\pi r dr}\right)\dfrac{\Delta t}{\tau_{100}}.
\end{equation}
We limit the formation of \embs to one \emb for each trap lifetime, and also restrict new \embs being formed within $0.5 H_{\rm gas}$ of the trap for the lifetime of the trap that formed the \emb.
This stops multiple \embs forming in extremely close proximity, where it would be unlikely for multiple \emb forming traps to arise simultaneously.

\begin{figure}
\centering
\includegraphics[scale=0.5]{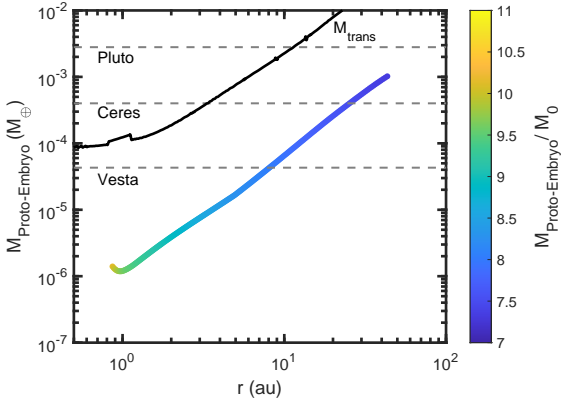}
\caption{The masses of \embs as a function or orbital distance that form in our fiducial disc model (initial disc mass of $0.1 \msun$). The colour code shows the \emb mass as a function of the characteristic planetesimal mass $M_0$ (eq. \ref{eq:MG_Liu}). The black line denotes the location of the transition mass (eq. \ref{eq:mtrans}).}
\label{fig:embryo_masses}
\end{figure}

Figure \ref{fig:embryo_masses} shows the masses and radial locations of the \embs formed in our fiducial model.
The colour code shows the \emb mass as a function of the characteristic mass (eq. \ref{eq:MG_Liu}), indicating that the largest formed objects are always around and order of magnitude larger than the characteristic mass, consistent with \citet{Abod19}.
Near the iceline, the \emb masses are around $10^{-6} \me$, much smaller than the terrestrial objects in the Solar System, as well as most large asteroids, e.g. Vesta.
Further out in the disc, more massive \embs are able to form, with Vesta mass objects forming at around 10 $\au$, and Ceres mass objects forming around 25 $\au$.

Interestingly the masses of all of these formed \embs are at least an order of magnitude lower than the pebble transition mass as shown by the black line in fig. \ref{fig:embryo_masses}, which pebble accretion models use for their initial planet mass, as it is the mass where pebble accretion switches from the Bondi regime to the Hill regime \citep{Bitsch15,Bitsch19,Coleman19}.
These masses are also much lower than the typical initial planet masses in planetesimal accretion models \citep[e.g.][]{Mordasini15,ColemanNelson14,ColemanNelson16,ColemanNelson16b}.
To obtain the masses used in these other works, the \embs would have to accrete either pebbles or local planetesimals, which we will examine in sect. \ref{sec:accretion}.

In our fiducial model, we find that the total initial mass of the \embs is equal to $0.02 \me$, $0.08\%$ of the total mass of planetesimals.
For the less massive disc, the total initial \emb mass was equal to $0.007 \me$ ($0.05\%$ of total planetesimal mass), whilst for the more massive disc, it was $0.07 \me$ ($0.13\%$ of total planetesimal mass). 
As can be expected the total mass in \embs is much smaller than that in the remaining planetesimals.

\section{Inclusion of pebble and planetesimal Accretion}
\label{sec:accretion}
Once \embs have formed in the disc, they begin to accrete the surrounding solid material.
This solid material can take the form of planetesimals formed through the streaming instability, or pebbles that are drifting past the \emb.
In this section we include the effects of planetesimal and pebble accretion individually and then combined to examine to what extent low-mass planets or giant planet cores are able to form.
Note here we do not include the effects of N-body interactions between \embs, planet migration or the accretion of gas, which will all be included in future work.
The aim here is to see whether the precursors to giant planets or planetary systems are able to form concurrently with the planetesimal and \emb formation process and whether there are preferential modes of accretion or regions where specific types of planets tend to form.

\subsection{Pebble Accretion}
\label{sec:acc_peb}

Since the pebbles in our simulations are contributing to the planetesimal formation process, the amount of pebbles drifting past the \embs is much reduced compared to previous works \citep[e.g.][]{Lambrechts14,Bitsch15,Coleman19}.
Still, as the remaining pebbles drift through the disc, they are able to be accreted more efficiently than planetesimals when passing through a \emb's Hill or Bondi sphere.
This is due to the increased gas drag forces that allowed them to become captured by the \emb's gravity \citep{Lambrechts12}.
To calculate this accretion rate, we follow \citet{Johansen17} by distinguishing between the Bondi regime (small bodies) and the Hill regime (massive bodies).
The Bondi accretion regime occurs for low mass bodies where they do not accrete all of the pebbles that pass through their Hill sphere, i.e. the body's Bondi radius is smaller than the Hill radius.
Once the Bondi radius becomes comparable to the Hill radius, the accretion rate becomes Hill sphere limited, and so the body accretes in the Hill regime.
Within our simulations, \embs typically begin accreting in the Bondi regime before transitioning to the Hill regime when they reach the transition mass,
\begin{equation}
\label{eq:mtrans}
    M_{\rm trans} \propto \eta^3 M_*.
\end{equation}
A further distinction within the two regimes, is whether the body is accreting in a 2D or a 3D mode.
This is dependent on the relation between the Hill radius of the body and the scale height of the pebbles in the disc.
For bodies with a Hill radius smaller than the scale height of pebbles, the accretion is in the 3D mode since pebbles are passing through the entire Hill sphere, whilst for bodies with a Hill radius larger than the pebble scale height, regions of the Hill sphere remain empty of pebbles and as such the accretion rate becomes 2D as the body's mass increases.
Following \citet{Johansen17} the equations for the 2D and 3D accretion rates are
\begin{equation}
\dot{M}_{\rm 2D} = 2 R_{\rm acc} \Sigma_{\rm peb} \delta v,
\end{equation}
and
\begin{equation}
\dot{M}_{\rm 3D} = \pi R_{\rm acc}^2 \rho_{\rm peb} \delta v,
\end{equation}
where $\Sigma_{\rm peb}$ is the pebble surface density, while $\rho_{\rm peb}$ is the midplane pebble density.
Here $\delta v = \Delta v + \Omega R_{\rm acc}$ is the approach speed, with $\Delta v$ being the sub-Keplerian velocity. 
The accretion radius $R_{\rm acc}$ depends on whether the accreting object is in the Hill or Bondi regime, and also on the friction time of the pebbles.
In order for pebbles to be accreted they must be able to significantly change direction on time-scales shorter than the friction time.
This inputs a dependence of the friction time onto the accretion radius, forming a criterion accretion radius $\hat{R}_{\rm acc}$ which is equal to 
\begin{equation}
 \hat{R}_{\rm acc} = \left( \frac{4 t_{\rm f}}{t_{\rm B}} \right)^{1/2} R_{\rm B},
\end{equation}
for the Bondi regime, and:
\begin{equation}
 \hat{R}_{\rm acc} = \left(  \frac{\Omega t_{\rm f}}{0.1} \right)^{1/3} R_{\rm H},
 \label{Racc_Hill}
\end{equation}
for the Hill regime.
Here $R_{\rm B}$ is the Bondi radius, while $R_{\rm H}$ is the Hill radius, $t_{\rm B}$ is the Bondi sphere crossing time, and $t_{\rm f}$ is the friction time
The accretion radius is then equal to
\begin{equation}
    R_{\rm acc} = \hat{R}_{\rm acc} \exp[-\chi(t_{\rm f}/t_{\rm p})^{\gamma}]
\end{equation}
where $t_{\rm p} = G M/(\Delta v + \Omega R_{H})^3$ is the characteristic passing time-scale, $\chi = 0.4$ and $\gamma = 0.65$ \citep{OrmelKlahr2010}.

The object then grows by accreting pebbles until it reaches the so-called pebble isolation mass, that is the mass required to perturb the gas pressure gradient in the disc: i.e. the gas velocity becomes super-Keplerian in a narrow ring outside the planet's orbit reversing the action of the gas drag.
The pebbles are therefore pushed outwards rather than inwards and accumulate at the outer edge of this ring stopping the core from accreting solids \citep{PaardekooperMellema06,Rice06}.
Initial works found that the pebble isolation mass was proportional to the cube of the local gas aspect ratio \citep{Lambrechts14}.
More recent work however has examined what effects disc viscosity and the stokes number of the pebbles have on the pebble isolation mass, finding that small pebbles that are well coupled to the gas are able to drift past the pressure bump exterior to the planet's orbit \citep{Ataiee18,Bitsch18}.
To account for the pebble isolation mass whilst including the effects of turbulence and stokes number, we follow \citet{Bitsch18}, and define a pebble isolation mass-to-star ratio,
\begin{equation}
q_{\rm iso} = \left(q_{\rm iso}^{\dagger} + \frac{\Pi_{\rm crit}}{\lambda} \right) \frac{M_{\oplus}}{M_*}
\end{equation}
where $q_{\rm iso}^\dagger  = 25 f_{\rm fit}$, and
\begin{equation}
\label{eq:ffit}
 f_{\rm fit} = \left[\frac{H/r}{0.05}\right]^3 \left[ 0.34 \left(\frac{\log(\alpha_3)}{\log(\alpha)}\right)^4 + 0.66 \right] \left[1-\frac{\frac{\partial\ln P}{\partial\ln r } +2.5}{6} \right] \ ,
\end{equation}
with $\alpha_3 = 0.001$.

\begin{figure}
\centering
\includegraphics[scale=0.5]{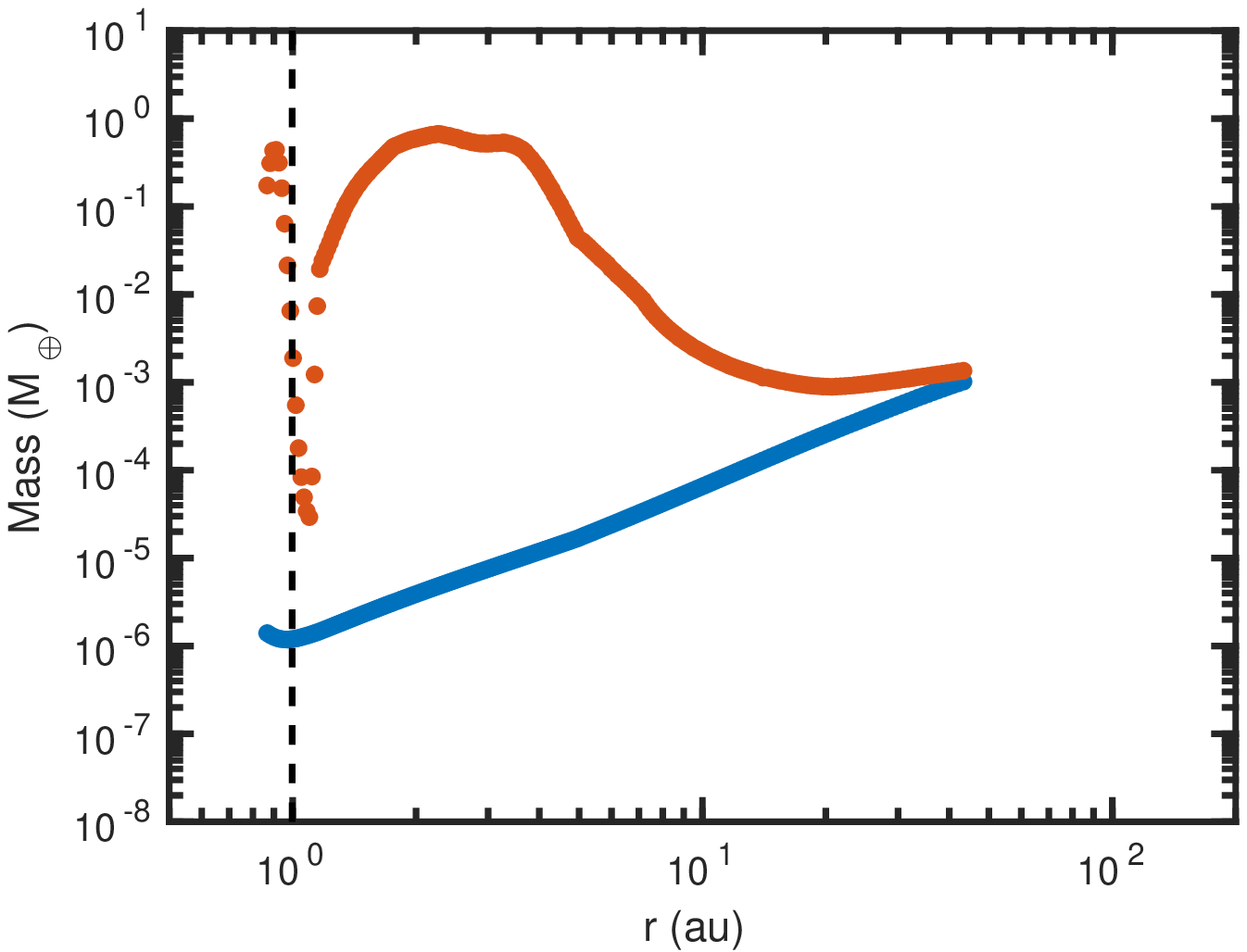}
\caption{Initial (blue) and final (red) masses of \embs that form in a $0.1 \msun$ protoplanetary disc and able to accrete pebbles. The vertical dashed line shows the location of the water iceline.}
\label{fig:pebble_masses}
\end{figure}

\begin{figure}
\centering
\includegraphics[scale=0.5]{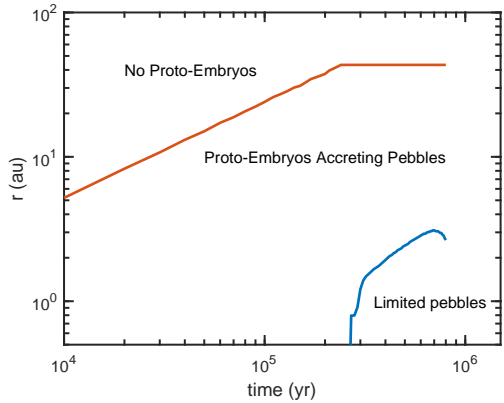}
\caption{The locations of the innermost (blue line) and outermost (red line) accreting \emb as a function of time. \embs orbiting at distances below the blue line are unable to accrete pebbles due to their supply of pebbles being accreted by more distant \embs. Above the red line, no \embs have been able to form at that time. In between the red and blue lines, \embs are able to accrete pebbles at
accretion rates detailed in sect. \ref{sec:acc_peb}.}
\label{fig:pebble_accretion_time}
\end{figure}

Figure \ref{fig:pebble_masses} shows the initial and final \emb mass as a function of orbital radius when incorporating eqs. \ref{eq:mtrans}--\ref{eq:ffit} into out fiducial model.
The blue points show the initial \emb masses, of which are identical to those in fig. \ref{fig:embryo_masses}.
The red points in fig. \ref{fig:pebble_masses} show the final \emb masses at the end of the disc lifetime after they have accreted pebbles flowing past their orbital position.
Close to the star, interior to and around the water iceline as shown by the vertical dashed line, pebble accretion appears very efficient.
This is due to slightly smaller pebbles drifting past the \embs here as the Stokes number becomes fragmentation limited, reducing the pebble size.
The reduction in pebble size reduces their radial velocity, allowing the pebble surface density to increase since there is then a small pile up pebbles, as can be seen by the increase in the purple line at $\sim 1\au$ in fig. \ref{fig:sigmas}.
This enhancement in the pebble surface density allows the \embs around the iceline to grow significantly, even though they are accreting in the inefficient bondi regime.
Slightly further out, where the Stokes number begins to be drift dominated, the pebbles are larger, and so there is no enhancement in the pebble surface density, and so accretion is very inefficient, as can be seen by the red points at around $\sim 10^{-4} \me$ just exterior to 1 $\au$ in fig. \ref{fig:pebble_masses}.

As the \embs form further out in the disc, their initial masses increase, to the point where they are able to accrete significant amounts of pebbles over time.
This accretion continues until the pebble supply is reduced to minimal levels, causing the accretion rate to again become inefficient. 
The reasons for the reduction in the pebble supply is a combination of  pebbles being trapped further out in the disc, ultimately forming more planetesimals and \embs, as well as these later formed \embs at larger separations from the star accreting pebbles, further reducing the pebble flux reaching the inner regions of the disc.
This is evident in fig. \ref{fig:pebble_accretion_time} where the red line shows the outer-most orbit that a \emb is accreting pebbles, whilst the blue line shows the location where the pebble flux through the disc has reduced to the 10$\%$ level.
Note that the $1\%$ level is only slightly closer to the star than the 10$\%$ level.
For \embs with $r< 2\au$, their ability to accrete pebbles is significantly reduced after $\sim0.4$ Myr, since the \embs located further out were accreting significant amounts of pebbles.
Even though their pebble supply was diminished, these \embs were able to reach terrestrial mass.
Interestingly, none of the \embs here were large enough to reach the pebble isolation mass.
In future work, N-body interactions between the \embs will be included which may allow some of these \embs to grow to even larger masses and reach the pebble isolation mass.
Reaching the pebble isolation mass will also restrict the mass growth of \embs closer to the star since the pebbles will become trapped at the pressure bump induced by the isolating \embs.

For \embs that formed at distances greater than 10 $\au$, these were unable to accrete significant amounts of pebbles.
Even though their initial masses were closer to their transition masses, the pebble midplane density is ever diminishing the further out in the disc the \embs try to accrete, reducing the number of interactions between pebbles and \embs over an orbit.
This results in the mass accretion rate gradually dropping with increasing orbital distance.
The other factor affecting these \embs is the further out in the disc they reside, the later they formed.
This gives them less time to accrete pebbles before the pebble production front reaches the outer edge of the disc, where the mass flux of pebbles drops to zero, of which this occurs after $\sim 0.9$ Myr.
This is particularly evident for those \embs that formed extremely far out in the disc at distances greater than $30 \au$, where their final masses are similar to their initial masses, being at most a factor two more massive.
This allows the more massive embryos to reach masses roughly two thirds the mass of Pluto.

So from the pebble accretion scenario, it is evident that there are preferential regions of the disc where it is favoured.
Around the water iceline, pebble accretion is efficient due to the smaller, slower moving, fragmentation limited pebbles allowing the \embs there to accrete more efficiently.
Slightly further out, where there is no surface density enhancement, accretion is inefficient due to \emb masses being significantly lower than the transition mass, and from the pebble supply being cut off after $\sim 0.3$ Myr.
In the outer regions of the disc the accretion time-scales are too long, and the time available for accretion is also reduced, resulting again in \embs close to their initial, albeit much more massive, mass.
Whilst in the middle region of the disc between the iceline and $\sim$few$\au$, pebble accretion can be quite efficient allowing multiple terrestrial mass \embs to grow, much more massive than they were at their formation, of which with further N-body interactions and gas accretion, could form into the more massive planets seen in recent observations.
In total, \embs accounted for a mass of $40.5 \me$, with the most massive \emb having a mass of $0.66 \me$.

\subsection{Planetesimal Accretion}
\label{sec:acc_pltml}

Whilst the section above examined the effects of pebble accretion on the formed \embs , we now examine what effects planetesimal accretion will have.
Since we are counting \embs as single bodies in the disc, we do not class planetesimals as super particles such as seen in previous works \citep[e.g.][]{ColemanNelson14,ColemanNelson16,ColemanNelson16b}.
Instead we follow other works that treat the accretion of planetesimals from an evolving surface density, essentially treating the planetesimals as a fluid-like disc \citep[e.g.][]{Alibert2006,Ida13,Mordasini15}.
We follow \citet{Fortier13} in calculating the planetesimal accretion rate that depends on the inclination and eccentricity evolution of the planetesimals.
The planetesimal surface density $\Sigma_{\rm pltml}$ evolves as planetesimals are accreted by \embs.
We also evolve the eccentricity rms $e_{\rm pltml}$ and inclination rms $i_{\rm pltml}$ by solving the differential equations for self-stirring \citep{Ohtsuki99}, gravitational stirring by nearby \embs \citep{Ohtsuki99} and also the effects of gas disc damping \citep{Adachi,Inaba01,Rafikov04}.

The planetesimal accretion rate is equal to
\begin{equation}
\label{eq:mdotpltml}
    \dot{M}_{\rm pltml} = \Omega \bar{\Sigma}_{\rm pltml} R_{\rm H}^2 P_{\rm coll}
\end{equation}
where $\bar{\Sigma}_{\rm pltml}$ is the average planetesimal surface density in the \embs' feeding zone (taken here to be equal to 10 Hill radii) and $P_{\rm coll}$ is the collision probability following \citet{Inaba01},
\begin{equation}
\label{eq:P_col}
P_{\rm coll}=\min\left( {P}_{\rm med},\left({P}^{-2}_{\rm high}+{P}^{-2}_{\rm low}\right)^{-1/2}\right)
\end{equation}
where the individual components are equal to
\begin{equation}
    P_{\rm high} = \frac{(r_{\rm emb}+r_{\rm pltml})^2}{2\pi R_{\rm H}^2}\left(I_{\rm F}(\beta)+\frac{6R_{\rm H}I_{\rm G}(\beta)}{(r_{\rm emb}+r_{\rm pltml})\tilde{e}^2} \right)
\end{equation}
\begin{equation}
    P_{\rm med} = \frac{(r_{\rm emb}+r_{\rm pltml})^2}{4\pi R_{\rm H}^2\tilde{i}}\left(17.3+\frac{232R_{\rm H}}{r_{\rm emb}+r_{\rm pltml}}\right)
\end{equation}
\begin{equation}
    P_{\rm low} = 11.3\left(\frac{r_{\rm emb}+r_{\rm pltml}}{R_{\rm H}}\right)^{1/2}
\end{equation}
where $P_{\rm high}$, $P_{\rm med}$ and $P_{\rm low}$ are the collision probabilities for different velocity regimes that depend on the random velocities of the planetesimals.
The quantities $\tilde{e}$ and $\tilde{i}$ are the reduced eccentricities ($\tilde{e}=ae/R_{\rm H}$) and inclinations ($\tilde{i}=ai/R_{\rm H}$).
These quantities indicate which velocity regime the planetesimals are found in depending on their relative velocities, with: the high-velocity regime for $\tilde{e},\tilde{i} \ge 2$, the medium velocity regime being for $2\ge\tilde{e},\tilde{i} \ge 0.2$ and the low-velocity regime for $\tilde{e},\tilde{i} < 0.2$.
The variable $\beta$ in the above equations is equal to $\tilde{i}/\tilde{e}$ and the functions $I_{\rm F}(\beta)$  $I_{\rm G}(\beta)$ are well approximated by
\begin{equation}
    I_{\rm F}(\beta) \backsimeq \frac{1+0.95925\beta+0.77251\beta^2}{\beta(0.13142+0.12295\beta)}
\end{equation}
\begin{equation}
    I_{\rm F}(\beta) \backsimeq \frac{1+0.3996\beta}{\beta(0.0369+0.048333\beta+0.006874\beta^2)}
\end{equation}
for $0\le\beta\le 1$, which is the range of $\beta$ values within this work \citep{Chambers06}.

Since the planetesimal eccentricities and inclinations have a large impact on the planetesimal accretion rate, it is necessary to define these values and allow them to evolve over the course of the disc lifetime.
Planetesimals experience gas drag from the disc which acts to damp the eccentricities whilst simultaneously experiencing gravitational interactions with \embs as well as gravitational interactions and minor collisions with fellow planetesimals.
In incorporating these processes into the evolution of the planetesimal eccentricities and inclinations, we obtain
\begin{equation}
    \dfrac{de^2}{dt}=\dfrac{de^2}{dt}\bigg|_{\rm drag}+ \dfrac{de^2}{dt}\bigg|_{\rm pltml} + \dfrac{de^2}{dt}\bigg|_{\rm emb}
\end{equation}
\begin{equation}
    \dfrac{di^2}{dt}=\dfrac{di^2}{dt}\bigg|_{\rm drag}+ \dfrac{di^2}{dt}\bigg|_{\rm pltml} + \dfrac{di^2}{dt}\bigg|_{\rm emb}
\end{equation}
where the subscripts `drag', `pltml' and `emb' refer to the contributions from gas drag damping, mutual stirring by planetesimals, and gravitational stirring by \embs.
For the calculation of the three terms affecting the eccentricity and inclination evolution, we follow \citet{Fortier13} where the equations and contributions of these terms can be found in their equations 31--53.

\begin{figure}
\centering
\includegraphics[scale=0.5]{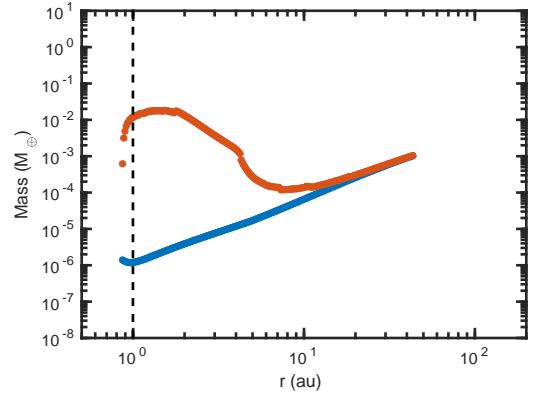}
\caption{Initial (blue) and final (red) masses of \embs that form in a $0.1 \msun$ protoplanetary disc and able to accrete planetesimals. The vertical dashed line shows the location of the water iceline.}
\label{fig:planetesimal_masses}
\end{figure}

\begin{figure}
\centering
\includegraphics[scale=0.5]{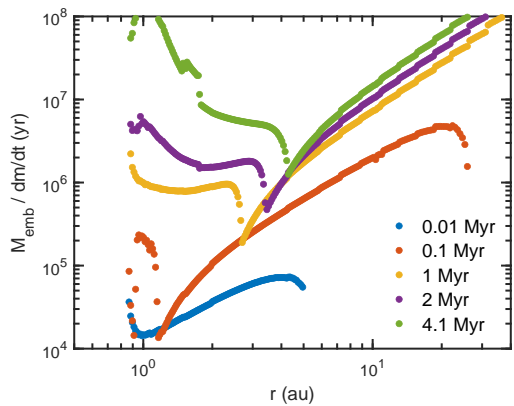}
\caption{Mass doubling time-scales for \embs accreting planetesimals in a $0.1 \msun$ protoplanetary disc at times of: 0.01 Myr (blue), 0.1 Myr (red), 1 Myr (yellow), 2 Myr (purple) and 4.1 Myr (green). The green points corresponding to 4.1 Myr, also correspond to the time-scales at the end of the disc lifetime.}
\label{fig:planetesimal_accretion_time}
\end{figure}

With the eccentricity and inclination now known for the planetesimals, we allow the \embs to accrete planetesimals using eq. \ref{eq:mdotpltml}, and remove the accreted mass from the \embs local planetesimal surface densities.
Again, using our fiducial disc setup, fig. \ref{fig:planetesimal_masses} shows the initial (blue points) and final (red points) masses of the \embs as a function of their orbital radius.
The initial \emb radii and masses are identical to those in figs. \ref{fig:characteristic_radius} and \ref{fig:embryo_masses} respectively, with both increasing as a function of orbital distance.

Looking at the red points in fig. \ref{fig:planetesimal_masses}, we can see that the majority of the planetesimal accretion experienced by the \embs occurs to those \embs close to the central star.
In these inner regions of the disc the planetesimals are still relatively small, $r_{\rm pltml} < 100$km, allowing for the more massive \embs to accrete in the runaway accretion regime before the planetesimal eccentricities and inclinations were significantly excited by the now more massive \embs.
Note we don't include the migration effects of gas drag acting on the planetesimals in these simulations, but given the large size of the planetesimals in the disc, gas drag should have a negligible effect \citep{Adachi,Weidenschilling_77}.
Shepherding of planetesimals through interactions with \embs is also neglected, but again, given the large size of the planetesimals and large relative velocities they obtain through interactions with \embs and other planetesimals, this would also be of negligible effect \citep{ColemanNelson16}.
The excitation of the planetesimal eccentricities occurs after only $\sim 0.3$ Myr in the region just exterior to the iceline reducing the accretion rate for the surrounding \embs and leaving them accreting very slowly whilst having masses, $m_{\rm emb} \sim 2\times10^{-3}\me$.

As the orbital radius of \embs increases, their accretion rates drop as the planetesimal eccentricities increase.
The planetesimal radii also increase with orbital radius, which greatly influences the efficiency of gas drag acting on the planetesimals in reducing their eccentricities \citep[see eqs. 31-41 of ][]{Fortier13}.
This allows the planetesimal eccentricities to maintain larger equilibrium values, hindering their accretion onto \embs.
This results in the final \emb masses gradually becoming smaller as a function of orbital distance and can be seen for all \embs exterior to $\sim1.5 \au$ and out to $\sim 4\au$ in our fiducial disc model, where the planetesimal eccentricities have been significantly excited reducing the accretion rate.
This leaves these \embs with masses between $10^{-3}$ to $10^{-2}\me$.
Around a mass of $10^{-3}\me$, the \embs begin to significantly increase the planetesimal eccentricities, forcing their equilibrium eccentricity (interactions balanced by gas drag) to increase, resulting in substantially lower accretion rates.
Up to to $4\au$, this becomes much more apparent since the planetesimals are much more easily excited to larger eccentricities, and as such the accretion rates drop very quickly once \embs reach this mass.
As can be seen in the outer disc ($r\ge 5 \au$) of fig. \ref{fig:planetesimal_masses}, the final \emb masses quickly drop to being similar to their initial masses (reaching their initial mass at around $10\au$), due to the very low accretion rates that arise from the easily excited planetesimal eccentricities.
Given the lack of planetesimal accretion in these outer regions, this can be a problem for forming giant planets, since when migration is included, the giant planet cores need to form and undergo runaway gas accretion at large orbital distances in order to survive as Cold Jupiters \citep[$r_{\rm p}>1\au$,][]{ColemanNelson14,ColemanNelson16b}.

Given the low accretion rates experienced by the \embs in the planetesimal accretion scenario, it is interesting to look at the time-scale required for an \emb to double its mass.
Figure \ref{fig:planetesimal_accretion_time} shows this mass doubling time-scale at a range of times in the disc.
Looking very early in the disc lifetime, after 0.01 Myr (blue points), the very short accretion time-scales of \embs close to the central star, around $1 \au$ can easily be seen.
For a period of time, these time-scales can be around $10^4$ years allowing for the \embs to very quickly increase in mass.
After 0.1 Myr (red points) these initially fast accretors, have excited the planetesimal eccentricities and inclinations, significantly increasing their accretion time-scales.
At this point, the faster accretors are located just exterior to the iceline around 1.5 $\au$.
Further out in the disc the accretion time-scales are at least $10^5$ yrs, with time-scales on order typical disc lifetimes ($>10^6$ yrs) for \embs exterior to 10 $\au$.
After 1 Myr (yellow points), the accretion time-scales for \embs around the iceline have now risen to $\sim10^6$ yrs, since the planetesimal eccentricities have become significantly excited, reducing the probability of impacts between the planetesimals and the \embs that now have a mass $\sim 10^{-2} \me$.
The `wave' of \embs accreting faster than their neighbours is now at the \embs at $\sim 3\au$, where their accretion time-scales are around $2\times 10^5$ yrs, still significantly long for allowing the \embs to accrete enough planetesimals to become giant planet cores.
The patterns described above continue for the purple and green points, showing the accretion time-scales at 2 and 4.1 Myr respectively.
The latter green points also correspond to the end of the disc lifetime, and the undamped conditions due to gas drag for the planetesimals.
It would be expected that the accretion time-scales would therefore increase after the end of the disc lifetime, since planetesimals would now only be excited by mutual interactions and by \emb interactions, with no balancing force to create an equilibrium.

Looking at the outer disc region in fig. \ref{fig:planetesimal_accretion_time}, exterior to $10 \au$, the accretion times are on the order of the disc lifetime indicating that very little accretion occurs for these \embs, as is shown when comparing the final and initial \emb masses for these \embs in fig. \ref{fig:planetesimal_masses}.
As highlighted above the reasons for such long accretion time-scales is due to the very large planetesimal eccentricities due to the easiness for dynamical interactions to excite them, as well as the large planetesimal sizes reducing the effects of gas drag.

Whilst the amount of planetesimal accretion here is limited, at most locations of the disc, it is worth noting that these simulations do not include N-body interactions between \embs.
These interactions could allow more massive cores to form that could increase the accretion rates.
However these cores would also increase the planetesimal eccentricities which would further reduce the accretion rates.
As such, the difficulties in forming giant planet cores, or even low mass planets in the outer regions of the disc, may still be extremely difficult with such large and easily excitable planetesimals even with the inclusion of N-body interactions.
Smaller planetesimal sizes would significantly enhance the accretion rate as they would circularise more easily, but that would also require significant grinding down of the planetesimals within the disc lifetime, which is especially unlikely in the outer regions of the disc.

\begin{figure}
\centering
\includegraphics[scale=0.5]{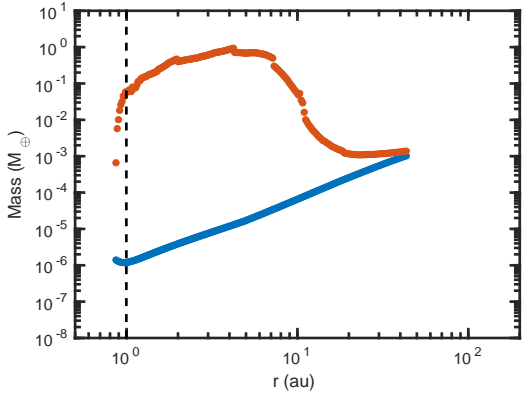}
\caption{Initial (blue) and final (red) masses of \embs that form in a $0.1 \msun$ protoplanetary disc and are able to accrete both pebbles and planetesimals. The vertical dashed line shows the location of the water iceline.}
\label{fig:combined_masses}
\end{figure}

\subsection{Combined Accretion}
\label{sec:acc_combined}

Where sects. \ref{sec:acc_peb} and \ref{sec:acc_pltml} examined the evolution of the \embs when accreting either pebble or planetesimals, we now repeat the simulations with the two accretion mechanisms working in tandem.
Figure \ref{fig:combined_masses} shows the final (red) and initial (blue) \emb masses as a function of their orbital radius, with the vertical dashed line showing the location of the water iceline.
For the \embs around the iceline, they have been able to grow to $m_{\rm emb}\sim 10^{-2}$--$10^{-1} \me$.
This accretion has been from a combination of both pebble and planetesimal accretion.
Like the \embs in the pebble accretion scenario, the mass growth of these \embs ceased when the flow of pebbles past their orbits was stopped, due to the pebbles being accreted by more distant \embs as well as by planetesimal formation events.

For \embs located further away from the iceline, these were able to grow very efficiently in this combined accretion scenario.
Through pebble accretion they were quickly able to reach masses of 10$^{-3}\me$, where once at, those close to the iceline could then effectively accrete planetesimals, further raising their mass.
This efficient combined accretion continued out to an orbital radius of $\sim 10\au$ for our fiducial disc.
For those \embs located at orbital radii greater than $10 \au$, the amount of solids accreted through planetesimal accretion was negligible compared to pebble accretion.
This was due to the \embs again stirring up the eccentricities of the large planetesimals, considerably reducing their planetesimal accretion rates.
As such the \emb masses located exterior to $10\au$ in this combined accretion scenario are very similar to those found in the pebble accretion scenario.
This comparison continues for \embs further out in the disc at large orbital radii ($>20\au$), where the planetesimal accretion becomes even more insignificant compared to that of pebble accretion.

From this combined accretion scenario, it therefore seems that planetesimal accretion is only effective around the water iceline and in those few $\au$ exterior.
Around the iceline, the planetesimal accretion supplements pebble accretion, whilst further out in the disc, planetesimal accretion aids the pebble accretion rates by allowing the \embs to reach masses more comparable to the transition mass, where pebble accretion becomes more efficient.
However for \embs further out in the disc, with orbital radii greater than $10\au$, pebble accretion would seem to be the dominant form of accretion, since the planetesimal accretion rate there is hindered substantially by the easily excitable planetesimal eccentricities once \embs begin to significantly stir them up.
Still though, with the initial \emb masses being much below that of the transition mass, pebble accretion in itself there, remains inefficient.

\subsection{Comparison of Accretion Models}
\label{sec:acc_comparison}

\begin{figure}
\centering
\includegraphics[scale=0.5]{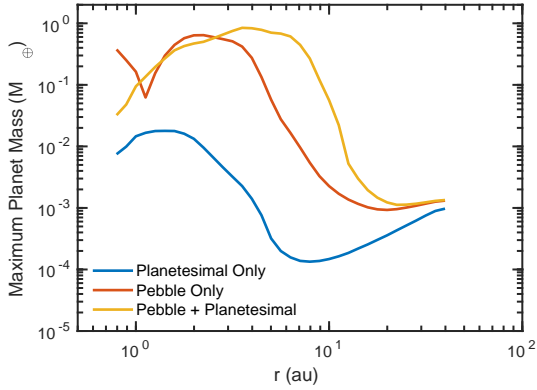}
\caption{The maximum \emb mass as a function of orbital radius for \embs accreting in different scenarios in a $0.1 \msun$ protoplanetary disc. The different scenarios correspond to: planetesimal only (blue line), pebble only (red line), both pebble and planetesimal (yellow line).}
\label{fig:max_planet_mass}
\end{figure}

\begin{figure}
\centering
\includegraphics[scale=0.5]{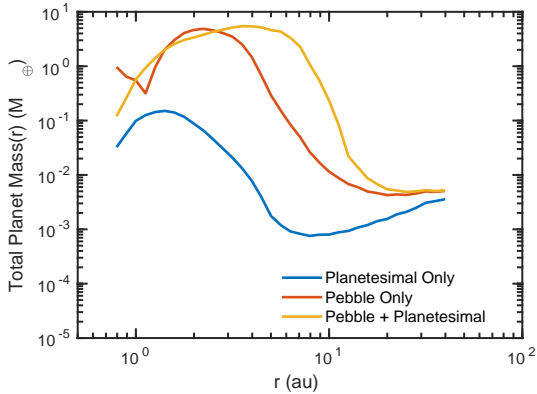}
\caption{The total \emb mass as a function of orbital radius for \embs accreting in different scenarios in a $0.1 \msun$ protoplanetary disc. Total masses are summed over an orbital radius of 0.05 dex. The different scenarios correspond to: planetesimal only (blue line), pebble only (red line), both pebble and planetesimal (yellow line).}
\label{fig:total_planet_mass}
\end{figure}

\subsubsection{Proto-Embryo Masses}

With the different accretion scenarios described above, it is clear that there is a significant impact on the types of planets that form from the initial \embs and their subsequent accretion.
In fig. \ref{fig:max_planet_mass} we compare the maximum \emb mass as a function of orbital radius for the different accretion routines.
When comparing the planetesimal (blue line) to the pebble (red line) accretion scenarios, it is clear that they give similar results in that their accretion is most efficient near and just exterior to the iceline.
Planetesimal accretion there is efficient due to the smaller planetesimals and short accretion time-scales, whilst consistent pebble supply and accretion over long times allows the \embs to significantly grow before the pebble supply is substantially reduced stopping the \embs from reaching the pebble isolation mass.
As the \embs form further out in the disc, out to a few $\au$, pebble accretion remains the dominant mode of accretion, where \emb masses are normally an order of magnitude larger in the pebble accretion scenario.
Exterior to $\sim$few $\au$, both modes of accretion reduce in their effectiveness.
Planetesimal accretion reduces due to increases in planetesimal eccentricities further out in the disc, whilst pebble accretion diminishes due to the lower pebble midplane densities and \embs remaining much smaller than the transition mass.
At large distances in the disc, both scenarios are relatively inefficient, albeit pebble accretion is slightly more efficient than planetesimal accretion, as the \embs are at most a factor few larger than their initial mass.

When combining the modes of accretion, it is clear that from the yellow line, that planetesimal accretion aids \emb growth further out in the disc for \embs accreting pebbles.
This can be seen by the extension of the most massive \emb being at $\sim 7\au$ in the combined accretion case compared to $\sim 4\au$ in the pebble accretion case.
With the \embs accreting both planetesimals and pebbles, they were able grow to the pebble transition mass through planetesimal accretion at larger orbital radii than when it was not included, where they were then able to more efficiently accrete pebbles.
However whilst planetesimal accretion aided \emb growth up to $\sim 10\au$, at larger distances the most massive \embs in the combined case are similar to those in the pebble case, showing that even with both accretion mechanisms functioning, mass growth at large orbital distance remains extremely difficult.
This leads to the maximum \emb masses at $\sim 40\au$ being roughly equal to 0.5--0.66 times the mass of Pluto, indicating that Pluto may not have significantly grown from its initial formation mass.

Whilst fig. \ref{fig:max_planet_mass} shows the maximum \emb mass for each accretion scenario, fig. \ref{fig:total_planet_mass} shows the total \emb mass as a function of orbital radius.
Looking at the total \emb masses is important as it allows us to examine how much mass there is in a specific area (i.e. around the iceline) that could then be concentrated in only a few bodies should N-body interactions be included.
To calculate the total mass, we sum the \emb masses over an orbital radius range of 0.05 dex (i.e. 0.05 orders of magnitude).
The profiles in fig. \ref{fig:total_planet_mass} are similar to those in fig. \ref{fig:max_planet_mass}, except for the masses are an order of magnitude higher due to the concentration of mass into only a few bodies.
Around the iceline, the \embs in the pebble and combined accretion scenarios are able to reach terrestrial masses, whilst those in the planetesimal accretion scenario remain at sub-Mars mass levels.

Further out in the disc, at a few $\au$, the total \emb mass in the planetesimal accretion scenario begins to quickly drop, with the most massive combined mass being less than $10^{-2}\me$ at orbital radii greater than $4\au$.
The main cause of this drop in accretion are the larger planetesimal eccentricities, easily excited by interactions with \embs , as well reductions in gas drag effectiveness due to large planetesimal sizes.
This leads to the total \emb mass there to approximately become similar to the initial total \emb mass at around 10 $\au$.
The planetesimal accretion results are in contrast to the pebble and combined accretion scenarios where the total \emb masses are in the super-Earth mass regime, out to $\sim 8 \au$ in the combined accretion scenario.
This is important as should these \embs reach these masses through N-body interactions, they can then begin to efficiently accrete gas and possibly become giant planets.
In total, between $2–-10 \au$ the \embs in the planetesimal accretion scenario totalled 0.54 $\me$, whilst the pebble and combined accretion scenarios totalled 34.7 $\me$ and 65.1 $\me$ respectively, highlighting the vast differences in accretion efficiencies between planetesimal and pebble accretion in this region.

For \embs exterior to $10 \au$, in the pebble and combined accretion scenarios, the total \emb mass reduced until around $20 \au$, where they then remained level at around $6\times 10^{-3} \me$.
This is due to the decreasing accretion efficiency of pebbles in this region, as well as the reduced amount of time that \embs can accrete for before the pebble production front reaches the outer edge of the disc.
For the planetesimal accretion scenario, the total \emb mass increases with orbital radius, but the final \emb masses are still less than a factor two greater than their initial masses.
In total, the summed \emb masses exterior to $10 \au$ totalled less than 0.2 $\me$ in all three scenarios, highlighting the lack of accretion at large orbital distances.

\subsubsection{N-body Interactions}

\begin{figure}
\centering
\includegraphics[scale=0.5]{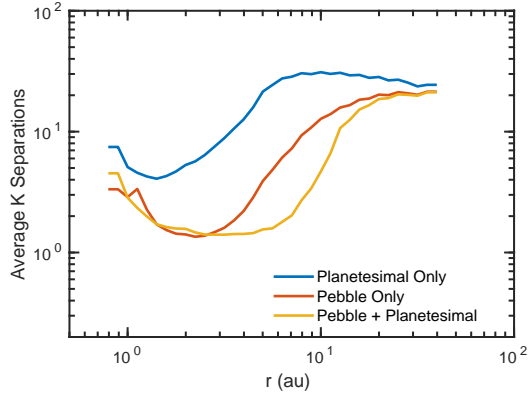}
\caption{The average mutual Hill separation (K) for the most massive \embs as a function of orbital distance. The different colours correspond to different accretion scenarios: planetesimal only (blue line), pebble only (red line), both pebble and planetesimal (yellow line).}
\label{fig:average_K}
\end{figure}

\begin{figure}
\centering
\includegraphics[scale=0.5]{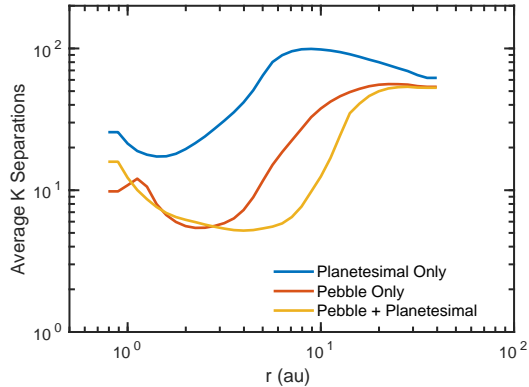}
\caption{The average mutual Hill separation (K) for the total \emb mass as a function of orbital distance, with \emb masses summed over 0.05 dex in orbital distance (i.e. 0.05 orders of magnitude). The different colours correspond to different accretion scenarios: planetesimal only (blue line), pebble only (red line), both pebble and planetesimal (yellow line).}
\label{fig:total_mass_K}
\end{figure}

As noted in the sections above, N-body interactions are not included in this work.
However from fig. \ref{fig:max_planet_mass}, numerous \embs of sufficient mass may be orbiting in extreme proximity where it might be indicated that N-body interactions could play an important role.
To examine this, fig. \ref{fig:average_K} shows the average mutual hill radius separation of the \embs at a given orbital radius.
Typically, \embs require a separation of at least 10 mutual hill radii to be stable for Gyr time-scales \citep{PuWu2015}.
As can be seen in fig. \ref{fig:average_K}, most of the \embs have mutual hill radii less than 10, with some cases only just greater than 1.
This implies that there would be significant dynamical evolution in these systems even when taking into account the presence of the gas disc, i.e. eccentricity and inclination damping.
For the region exterior to the iceline out to $\sim 10 \au$, the increase in \emb masses here would be extremely useful for the formation of giant planet cores.
Should N-body interactions allow these individual $\sim$Mars--Earth-mass \embs to grow into super-Earths as assumed to be the case for the total \emb masses seen in fig. \ref{fig:total_planet_mass}, then they could begin to accrete gas extremely quickly and become gas giants.
Whilst the inclusion of N-body interactions would likely have a significant impact on the \embs in the inner $10 \au$ of the disc, their impact on the more distant \embs would be less notable.
This is due to there being fewer \embs in the region, and the \embs being slightly more dynamically separated due to their low mass (with an average mutual hill radius of $\sim 2$).
As fig. \ref{fig:total_planet_mass} shows, even if these \embs did collide and form a more massive object, their masses would only reach $\sim$ Lunar-mass, far too insufficient for giant planets to form.
Depending on how early these interactions occur, and if these more massive \embs form, their pebble accretion rates would be enhanced until the pebble production front reaches the outer edge of disc, which would allow them to become more massive.

Whilst fig. \ref{fig:average_K} looks at the average separation of all of the \embs in the disc, fig. \ref{fig:total_mass_K} gives the mutual hill radius separation for the \embs found in fig. \ref{fig:total_planet_mass}.
Here it is assumed that the total mass of \embs for an orbital radius region of 0.05 dex have been assembled into a single \emb, i.e. have undergone numerous N-body collisions.
Figure \ref{fig:total_mass_K} then just examines the separation between these more massive \embs to examine whether fig. \ref{fig:total_planet_mass} gives an adequate representation of the final \emb masses as a function of orbital radius, without the inclusion of N-body interaction, migration or gas accretion.
For the planetesimal accretion scenario, it could be considered that the assumption has exaggerated the final \emb mass, as the minimum hill separation here is approximately 15, implying very stable systems.
Given that the maximum \emb mass here is $\sim 0.2 \me$, this would indicate significant problems for the planetesimal accretion scenario in forming planets with planetesimals of the size formed in this work, as there would be limited opportunities for these \embs to further increase their mass.

For the pebble and combined accretion scenarios, the total \emb masses appears adequate in estimating the \embs masses if N-body interactions were taken into account.
For most of the inner disc, where \embs have been able to grow, the average separation is between 5--15 mutual Hill radii, which can be stable for significant fractions of the disc lifetime, possibly going unstable after the gas disc has dissipated, with N-body interactions proceeding in undamped environments \citep{ColemanNelson16}.
If these \embs were able to form mean-motion resonances with their neighbours, then these orbits could be stable for time-scales much longer than the disc lifetime, whilst further interactions, or destabilising of resonant configurations, would most likely lead to further collisions, and an increase in \emb masses.
Like the planetesimal accretion scenario, the \embs in the outer regions of the disc ($>10\au$) the final \emb mass is most likely exaggerated since the average separations are larger than 50.
Even when looking at this region in fig. \ref{fig:average_K}, the \emb are at least 20 mutual hill radii apart, indicating that there would be very few interactions and collisions after the end of the disc lifetime.

Overall, from fig. \ref{fig:total_mass_K} it can be concluded that the \embs seen in fig \ref{fig:total_planet_mass} could be a good estimation of those formed in simulations that include N-body interaction.
Further complications that arise from planet migration and gas accretion would further influence the evolution of these \embs and along with N-body interactions will be examined in future work.

\section{Discussions and conclusions}
\label{sec:conclusions}

\subsection{Planetesimal and Proto-Embryo Formation}
In this paper we have explored the formation of planetesimals and \embs in evolving protoplanetary discs.
We model the planetesimal formation following \citet{Lenz19} where pebbles drifting through the protoplanetary can become trapped by short-lived local pressure bumps.
Should significant pebbles be trapped in these short-lived pressure bumps such that the local midplane dust-to-gas ratio exceeds unity, the streaming instability further concentrates the particles into filaments that might induce gravitational collapse.
Following \citet{Liu20}, the formed planetesimals have a specific characteristic radius, depending on the local disc properties.
With our fiducial disc model, containing a mass $10\%$ that of the central $1\msun$ star, we find $27 \me$ of planetesimals are formed throughout the disc lifetime extending from 0.8--45$\au$.
The planetesimal surface densities and overall masses are in agreement with other works \citep[e.g.][]{Lenz19,Voelkel20}, which include a considerably more intricate dust and pebble evolution model than that considered here.

In terms of the radius of the planetesimals, following \citep{Liu20} we find that the planetesimal radius increases with orbital distance.
Close to the central star, around the iceline, the planetesimals have a radius of $\sim$40 km.
However this radius increases considerably the further out in the disc the planetesimals form, with $r_{\rm pltml}\sim100$km at 4 $\au$ and  $r_{\rm pltml}\sim450$km at 40 $\au$.
This is seen to have a large impact on the planetesimal accretion rates, since the larger planetesimals experience weaker gas drag forces allowing them to retain larger eccentricities, that increases the relative velocities between \embs and planetesimals, reducing planetesimal accretion rates.
This is also different to what is assumed in other works \citep{Lenz19,Voelkel20} where they assume that the planetesimals have a size of 100 km ($r_{\rm pltml}=50$km) at every location in the disc.
The notion that planetesimals have a size of 100 km arises from studies of the main asteroid belt in the Solar System \citep{Morbidelli09,Delbo17}, whereas we assume the average planetesimal size depends on the local disc properties \citep{Liu20}.
Indeed, at the location of the main asteroid belt, the planetesimals formed in our fiducial disc have sizes between 60--100 km.

For \embs that form in our disc models, we assume that they are equal to the largest single planetesimal that forms in a planetesimal forming event.
With the \embs forming at more similar masses to the planetesimals, typically an order of magnitude greater in mass, like the planetesimals they therefore increase in mass as a function of orbital radius.
Close to the central star, the \embs form with masses $m_{\rm emb}\sim 10^{-6}\me$, but as the \embs form at larger orbital distances, their initial masses increase, reaching $m_{\rm emb}\sim 10^{-4}\me$ at 12$\au$ and $m_{\rm emb}\sim 10^{-3}\me$ at 40$\au$.
These masses are still below the initial embryo mass in previous works concerning planetesimal accretion \citep[e.g.][]{Mords09,ColemanNelson14,Mordasini15,ColemanNelson16,ColemanNelson16b} as well as the transition mass for the pebble accretion scenario \citep[e.g.][]{Bitsch15,Lambrechts19}, and as such significant accretion would be required for the \embs to grow to the initial embryo masses in such works.

\subsection{Accretion Scenarios}

In this paper, we also include accretion scenarios for \embs that form in the disc.
Once the \embs form, we allow them to accrete either pebbles or planetesimals, as well as being able to accrete both pebbles and planetesimals in a joint manner.\\
\noindent {\it Planetesimal Accretion}: Here we allowed \embs to accrete planetesimals by treating the planetesimals in a fluid-like manner following \citet{Fortier13}.
\embs in and around the water iceline were able to significantly increase their masses reaching masses $m_{\rm emb}\sim 10^{-2}\me$ before their accretion rates dropped due to the depletion of their feeding zones or planetesimal eccentricities became considerably excited.
Further out in the disc at $r>10\au$, the planetesimal accretion rates were relatively meagre meaning that few planetesimals were accreted, leaving the final \emb masses similar to their initial values.
This was due to the large planetesimal eccentricities, that are easily excitable, and weakly suppressed due to weak gas drag forces that arise with larger planetesimal sizes.
These results are in contrast to many other planetesimal accretion scenarios \citep[e.g.][]{Alibert2006,Ida13,Mordasini15,ColemanNelson14,ColemanNelson16,ColemanNelson16b}. where simulated planets have properties compatible to those observed.
However, these other works utilised much smaller planetesimals, ($r_{\rm pltml}<1$km in some cases) as well as much more massive initial \emb masses, both of which will act to considerably increase planetesimal accretion rates.\\
\noindent {\it Pebble Accretion}: We utilise the accretion formulae in \cite{Johansen17} to calculate the pebble accretion rates.
For \embs at the iceline, even though their initial mass is much smaller than the transition mass, their accretion rates are enhanced due to the pebble sizes being fragmentation limited.
With fragmentation limited pebbles, this results in enhanced pebble surface densities since the smaller pebbles drift inwards on time-scales more comparable to the gas.
Exterior to the iceline, where the pebble sizes are drift dominated, growth rates are small due to the larger pebbles and \embs accreting in the very inefficient bondi regime, before the pebble supply becomes depleted.
Further out in the disc, up to around 4 $\au$, the pebble accretion rate was fairly effective.
Numerous Mars-Earth mass \embs were able to form, with their accretion not being stopped through reaching the pebble isolation mass, but from a lack of pebbles in the disc.
This lack of accretable material arose from the pebbles being accreted by more distant \embs, limiting the supply from those closer to the central star.
Still numerous \embs were able to accrete appreciable amounts of pebbles and with the inclusion of N-body interactions, a number of giant planet cores should be able to form.
At larger distances, $r>10\au$, pebble accretion was much less effective since the \emb initial masses were much smaller than the transition mass as well as reduction in pebble midplane density further out in the disc.
They also had limited time to accrete pebbles before the pebble production front reached the outer edge of the disc, ceasing the generation of pebbles.\\
\noindent {\it Combined Accretion}: In combining the two accretion scenarios, \embs were able to accrete efficiently over a much larger region of the disc.
Whilst in the pebble accretion scenario, efficient accretion occurred up to $\sim 4\au$, in the combined accretion scenario, planetesimal accretion onto the initially low mass \embs allowed them to accrete pebbles more efficiently out to $\sim 8\au$.
This overall enhancement in accretion, allowed numerous Earth-mass \embs to form, which with the inclusion of N-body interactions could form into giant planet cores.
At larger distances in the disc, the role of planetesimal accretion quickly diminished as planetesimal eccentricities rose, greatly reducing the accretion rate.
This led to \emb masses being similar to those found in the pebble accretion scenario and remaining at around $\sim$Pluto-mass.

The results of these accretion scenarios are important when comparing to the planets or the expected precursors of planets observed today.
All scenarios were able to form numerous \embs that could be considered the precursors to the terrestrial planet formation scenarios \citep{Raymond14}.
The pebble and combined accretion scenarios were able to form a number of \embs that with the inclusion of N-body interactions, could be expected to merge into giant planet cores on orbits out to $\sim 4\au$ and $\sim 8\au$ respectively.
However neither scenarios were able to form \embs of significant mass at orbital radii greater than $10\au$.
Whilst giant planet cores could form at orbital radii closer than $10\au$, planet migration may be too effective in driving those cores in towards the central star.
Though this would be adequate for form hot Jupiters, this would be ineffective at forming cold Jupiters that need to form further out in the disc in order to survive migration processes \citep{ColemanNelson14,Bitsch15,ColemanNelson16b}.

\subsection{Future Work}

One significant drawback in the accretion scenarios in this work is the lack of N-body interactions.
In looking at the mutual hill separations between \embs (figs. \ref{fig:average_K} and \ref{fig:total_mass_K}), it is clear that considerable dynamical evolution would occur if N-body interactions were included.
In future work we will merge the models presented here with the \textsc{Mercury6} N-body integrator \citep{Chambers} similar to what has been achieved in recent works \citep[e.g.][]{ColemanNelson14,ColemanNelson16,ColemanNelson16b,Marleau19,Coleman19}.
With the inclusions of N-body interactions, \embs would no longer be on circular, coplanar orbits and as such, the effects of their eccentricities and inclinations would have to be included on the pebble and planetesimal accretion rates \citep{Liu18,Ormel18}.
We will also include prescriptions for planet migration, in both the type I regime when planets are embedded in the disc \citep{pdk10,pdk11} and the type II regime when planets have opened a gap in the disc \citep{LinPapaloizou86}.
In order to form giant planets we will also have to allow the \embs to accrete gas from the local disc.
In order to do this we will include new fits to the 1D envelope structure model of \citet{CPN17} that take into account the mass properties of the \emb, the \emb's local disc properties as well as opacity reduction factors \citep{Poon21}.

With most previous works examining the formation of planetary systems through either pebble or planetesimal accretion, by including the new prescriptions detailed above, along with the planetesimal and \emb formation models shown in this paper, will allow for a more complete beginning to end simulation for planets and planetary systems.
Only then will the effects of changing initial models and parameters such as the planetesimal formation efficiency \citep{Lenz20}, or the local environment \citep{Sellek20,ColemanHaworth20}, will we be able to determine where efforts need to lie in order to further explain the observed exoplanet populations.

\section*{Data Availability}
The data underlying this article will be shared on reasonable request to the corresponding author.

\section*{Acknowledgements}
GALC thanks the anonymous referee for their useful comments that improved the manuscript.
We also thank Craig Agnor, Giulia Ballabio, Remo Burn, Thomas Haworth and Christoph Mordasini for insightful discussions on the topics covered in this paper.
GALC was funded by the Leverhulme Trust through grant RPG-2018-418.

\bibliographystyle{mnras}
\bibliography{references}{}

\begin{thebibliography}{}
\makeatletter
\relax
\def\mn@urlcharsother{\let\do\@makeother \do\$\do\&\do\#\do\^\do\_\do\%\do\~}
\def\mn@doi{\begingroup\mn@urlcharsother \@ifnextchar [ {\mn@doi@}
  {\mn@doi@[]}}
\def\mn@doi@[#1]#2{\def\@tempa{#1}\ifx\@tempa\@empty \href
  {http://dx.doi.org/#2} {doi:#2}\else \href {http://dx.doi.org/#2} {#1}\fi
  \endgroup}
\def\mn@eprint#1#2{\mn@eprint@#1:#2::\@nil}
\def\mn@eprint@arXiv#1{\href {http://arxiv.org/abs/#1} {{\tt arXiv:#1}}}
\def\mn@eprint@dblp#1{\href {http://dblp.uni-trier.de/rec/bibtex/#1.xml}
  {dblp:#1}}
\def\mn@eprint@#1:#2:#3:#4\@nil{\def\@tempa {#1}\def\@tempb {#2}\def\@tempc
  {#3}\ifx \@tempc \@empty \let \@tempc \@tempb \let \@tempb \@tempa \fi \ifx
  \@tempb \@empty \def\@tempb {arXiv}\fi \@ifundefined
  {mn@eprint@\@tempb}{\@tempb:\@tempc}{\expandafter \expandafter \csname
  mn@eprint@\@tempb\endcsname \expandafter{\@tempc}}}

\bibitem[\protect\citeauthoryear{{Abod}, {Simon}, {Li}, {Armitage}, {Youdin}
  \& {Kretke}}{{Abod} et~al.}{2019}]{Abod19}
{Abod} C.~P.,  {Simon} J.~B.,  {Li} R.,  {Armitage} P.~J.,  {Youdin} A.~N.,
  {Kretke} K.~A.,  2019, \mn@doi [\apj] {10.3847/1538-4357/ab40a3}, \href
  {https://ui.adsabs.harvard.edu/abs/2019ApJ...883..192A} {883, 192}

\bibitem[\protect\citeauthoryear{{Adachi}, {Hayashi}  \& {Nakazawa}}{{Adachi}
  et~al.}{1976}]{Adachi}
{Adachi} I.,  {Hayashi} C.,   {Nakazawa} K.,  1976, \mn@doi [Progress of
  Theoretical Physics] {10.1143/PTP.56.1756}, \href
  {http://adsabs.harvard.edu/abs/1976PThPh..56.1756A} {56, 1756}

\bibitem[\protect\citeauthoryear{{Alexander} \& {Armitage}}{{Alexander} \&
  {Armitage}}{2007}]{Alexander07}
{Alexander} R.~D.,  {Armitage} P.~J.,  2007, \mn@doi [\mnras]
  {10.1111/j.1365-2966.2006.11341.x}, \href
  {http://adsabs.harvard.edu/abs/2007MNRAS.375..500A} {375, 500}

\bibitem[\protect\citeauthoryear{{Alexander} \& {Armitage}}{{Alexander} \&
  {Armitage}}{2009}]{Alexander09}
{Alexander} R.~D.,  {Armitage} P.~J.,  2009, \mn@doi [\apj]
  {10.1088/0004-637X/704/2/989}, \href
  {http://adsabs.harvard.edu/abs/2009ApJ...704..989A} {704, 989}

\bibitem[\protect\citeauthoryear{{Alexander} \& {Pascucci}}{{Alexander} \&
  {Pascucci}}{2012}]{AlexanderPascucci12}
{Alexander} R.~D.,  {Pascucci} I.,  2012, \mn@doi [\mnras]
  {10.1111/j.1745-3933.2012.01243.x}, \href
  {http://adsabs.harvard.edu/abs/2012MNRAS.422L..82A} {422, 82}

\bibitem[\protect\citeauthoryear{{Alibert} et~al.,}{{Alibert}
  et~al.}{2006}]{Alibert2006}
{Alibert} Y.,  et~al., 2006, \mn@doi [\aap] {10.1051/0004-6361:20065697}, \href
  {http://adsabs.harvard.edu/abs/2006A%26A...455L..25A} {455, L25}

\bibitem[\protect\citeauthoryear{{Anglada-Escud{\'e}}
  et~al.,}{{Anglada-Escud{\'e}} et~al.}{2016}]{Anglada2016}
{Anglada-Escud{\'e}} G.,  et~al., 2016, \mn@doi [\nat] {10.1038/nature19106},
  \href {http://adsabs.harvard.edu/abs/2016Natur.536..437A} {536, 437}

\bibitem[\protect\citeauthoryear{{Ataiee}, {Baruteau}, {Alibert}  \&
  {Benz}}{{Ataiee} et~al.}{2018}]{Ataiee18}
{Ataiee} S.,  {Baruteau} C.,  {Alibert} Y.,   {Benz} W.,  2018, \mn@doi [\aap]
  {10.1051/0004-6361/201732026}, \href
  {https://ui.adsabs.harvard.edu/abs/2018A&A...615A.110A} {615, A110}

\bibitem[\protect\citeauthoryear{{Bai} \& {Stone}}{{Bai} \&
  {Stone}}{2010}]{Bai10}
{Bai} X.-N.,  {Stone} J.~M.,  2010, \mn@doi [\apj]
  {10.1088/0004-637X/722/2/1437}, \href
  {https://ui.adsabs.harvard.edu/abs/2010ApJ...722.1437B} {722, 1437}

\bibitem[\protect\citeauthoryear{{Bai} \& {Stone}}{{Bai} \&
  {Stone}}{2014}]{Bai2014}
{Bai} X.-N.,  {Stone} J.~M.,  2014, \mn@doi [\apj]
  {10.1088/0004-637X/796/1/31}, \href
  {http://adsabs.harvard.edu/abs/2014ApJ...796...31B} {796, 31}

\bibitem[\protect\citeauthoryear{{Bell} \& {Lin}}{{Bell} \&
  {Lin}}{1994}]{Bell94}
{Bell} K.~R.,  {Lin} D.~N.~C.,  1994, \mn@doi [\apj] {10.1086/174206}, \href
  {http://adsabs.harvard.edu/abs/1994ApJ...427..987B} {427, 987}

\bibitem[\protect\citeauthoryear{{Bell}, {Cassen}, {Klahr}  \&
  {Henning}}{{Bell} et~al.}{1997}]{Bell97}
{Bell} K.~R.,  {Cassen} P.~M.,  {Klahr} H.~H.,   {Henning} T.,  1997, \apj,
  \href {http://adsabs.harvard.edu/abs/1997ApJ...486..372B} {486, 372}

\bibitem[\protect\citeauthoryear{{B{\'e}thune}, {Lesur}  \&
  {Ferreira}}{{B{\'e}thune} et~al.}{2016}]{BethuneLesur2016}
{B{\'e}thune} W.,  {Lesur} G.,   {Ferreira} J.,  2016, \mn@doi [\aap]
  {10.1051/0004-6361/201527874}, \href
  {http://adsabs.harvard.edu/abs/2016A%26A...589A..87B} {589, A87}

\bibitem[\protect\citeauthoryear{{Birnstiel}, {Dullemond}  \&
  {Brauer}}{{Birnstiel} et~al.}{2010}]{Birnstiel10}
{Birnstiel} T.,  {Dullemond} C.~P.,   {Brauer} F.,  2010, \mn@doi [\aap]
  {10.1051/0004-6361/200913731}, \href
  {https://ui.adsabs.harvard.edu/abs/2010A&A...513A..79B} {513, A79}

\bibitem[\protect\citeauthoryear{{Birnstiel}, {Klahr}  \&
  {Ercolano}}{{Birnstiel} et~al.}{2012}]{Birnstiel12}
{Birnstiel} T.,  {Klahr} H.,   {Ercolano} B.,  2012, \mn@doi [\aap]
  {10.1051/0004-6361/201118136}, \href
  {https://ui.adsabs.harvard.edu/abs/2012A&A...539A.148B} {539, A148}

\bibitem[\protect\citeauthoryear{{Bitsch}, {Lambrechts}  \&
  {Johansen}}{{Bitsch} et~al.}{2015}]{Bitsch15}
{Bitsch} B.,  {Lambrechts} M.,   {Johansen} A.,  2015, \mn@doi [\aap]
  {10.1051/0004-6361/201526463}, \href
  {http://adsabs.harvard.edu/abs/2015A%26A...582A.112B} {582, A112}

\bibitem[\protect\citeauthoryear{{Bitsch}, {Morbidelli}, {Johansen}, {Lega},
  {Lambrechts}  \& {Crida}}{{Bitsch} et~al.}{2018}]{Bitsch18}
{Bitsch} B.,  {Morbidelli} A.,  {Johansen} A.,  {Lega} E.,  {Lambrechts} M.,
  {Crida} A.,  2018, \mn@doi [\aap] {10.1051/0004-6361/201731931}, \href
  {https://ui.adsabs.harvard.edu/abs/2018A&A...612A..30B} {612, A30}

\bibitem[\protect\citeauthoryear{{Bitsch}, {Raymond}  \& {Izidoro}}{{Bitsch}
  et~al.}{2019}]{Bitsch19}
{Bitsch} B.,  {Raymond} S.~N.,   {Izidoro} A.,  2019, \mn@doi [\aap]
  {10.1051/0004-6361/201935007}, \href
  {https://ui.adsabs.harvard.edu/abs/2019A&A...624A.109B} {624, A109}

\bibitem[\protect\citeauthoryear{{Br{\"u}gger}, {Burn}, {Coleman}, {Alibert}
  \& {Benz}}{{Br{\"u}gger} et~al.}{2020}]{Brugger20}
{Br{\"u}gger} N.,  {Burn} R.,  {Coleman} G.~A.~L.,  {Alibert} Y.,   {Benz} W.,
  2020, \mn@doi [\aap] {10.1051/0004-6361/202038042}, \href
  {https://ui.adsabs.harvard.edu/abs/2020A&A...640A..21B} {640, A21}

\bibitem[\protect\citeauthoryear{{Carrera}, {Gorti}, {Johansen}  \&
  {Davies}}{{Carrera} et~al.}{2017}]{Carrera17}
{Carrera} D.,  {Gorti} U.,  {Johansen} A.,   {Davies} M.~B.,  2017, \mn@doi
  [\apj] {10.3847/1538-4357/aa6932}, \href
  {https://ui.adsabs.harvard.edu/abs/2017ApJ...839...16C} {839, 16}

\bibitem[\protect\citeauthoryear{{Carrera}, {Simon}, {Li}, {Kretke}  \&
  {Klahr}}{{Carrera} et~al.}{2021}]{Carrera21}
{Carrera} D.,  {Simon} J.~B.,  {Li} R.,  {Kretke} K.~A.,   {Klahr} H.,  2021,
  \mn@doi [\aj] {10.3847/1538-3881/abd4d9}, \href
  {https://ui.adsabs.harvard.edu/abs/2021AJ....161...96C} {161, 96}

\bibitem[\protect\citeauthoryear{{Chambers}}{{Chambers}}{1999}]{Chambers}
{Chambers} J.~E.,  1999, \mn@doi [\mnras] {10.1046/j.1365-8711.1999.02379.x},
  \href {http://adsabs.harvard.edu/abs/1999MNRAS.304..793C} {304, 793}

\bibitem[\protect\citeauthoryear{{Chambers}}{{Chambers}}{2006}]{Chambers06}
{Chambers} J.,  2006, \mn@doi [\icarus] {10.1016/j.icarus.2005.10.017}, \href
  {https://ui.adsabs.harvard.edu/abs/2006Icar..180..496C} {180, 496}

\bibitem[\protect\citeauthoryear{{Clarke}, {Gendrin}  \& {Sotomayor}}{{Clarke}
  et~al.}{2001}]{Clarke2001}
{Clarke} C.~J.,  {Gendrin} A.,   {Sotomayor} M.,  2001, \mn@doi [\mnras]
  {10.1046/j.1365-8711.2001.04891.x}, \href
  {http://adsabs.harvard.edu/abs/2001MNRAS.328..485C} {328, 485}

\bibitem[\protect\citeauthoryear{{Coleman} \& {Haworth}}{{Coleman} \&
  {Haworth}}{2020}]{ColemanHaworth20}
{Coleman} G. A.~L.,  {Haworth} T.~J.,  2020, \mn@doi [\mnras]
  {10.1093/mnrasl/slaa098}, \href
  {https://ui.adsabs.harvard.edu/abs/2020MNRAS.496L.111C} {496, L111}

\bibitem[\protect\citeauthoryear{{Coleman} \& {Nelson}}{{Coleman} \&
  {Nelson}}{2014}]{ColemanNelson14}
{Coleman} G.~A.~L.,  {Nelson} R.~P.,  2014, \mn@doi [\mnras]
  {10.1093/mnras/stu1715}, \href
  {http://adsabs.harvard.edu/abs/2014MNRAS.445..479C} {445, 479}

\bibitem[\protect\citeauthoryear{{Coleman} \& {Nelson}}{{Coleman} \&
  {Nelson}}{2016a}]{ColemanNelson16}
{Coleman} G.~A.~L.,  {Nelson} R.~P.,  2016a, \mn@doi [\mnras]
  {10.1093/mnras/stw149}, \href
  {http://adsabs.harvard.edu/abs/2016MNRAS.457.2480C} {457, 2480}

\bibitem[\protect\citeauthoryear{{Coleman} \& {Nelson}}{{Coleman} \&
  {Nelson}}{2016b}]{ColemanNelson16b}
{Coleman} G.~A.~L.,  {Nelson} R.~P.,  2016b, \mn@doi [\mnras]
  {10.1093/mnras/stw1177}, \href
  {http://adsabs.harvard.edu/abs/2016MNRAS.460.2779C} {460, 2779}

\bibitem[\protect\citeauthoryear{{Coleman}, {Nelson}, {Paardekooper},
  {Dreizler}, {Giesers}  \& {Anglada-Escud{\'e}}}{{Coleman}
  et~al.}{2017a}]{ColemanProxima17}
{Coleman} G.~A.~L.,  {Nelson} R.~P.,  {Paardekooper} S.~J.,  {Dreizler} S.,
  {Giesers} B.,   {Anglada-Escud{\'e}} G.,  2017a, \mn@doi [\mnras]
  {10.1093/mnras/stx169}, \href
  {http://adsabs.harvard.edu/abs/2017MNRAS.467..996C} {467, 996}

\bibitem[\protect\citeauthoryear{{Coleman}, {Papaloizou}  \&
  {Nelson}}{{Coleman} et~al.}{2017b}]{CPN17}
{Coleman} G.~A.~L.,  {Papaloizou} J.~C.~B.,   {Nelson} R.~P.,  2017b, \mn@doi
  [\mnras] {10.1093/mnras/stx1297}, \href
  {http://adsabs.harvard.edu/abs/2017MNRAS.470.3206C} {470, 3206}

\bibitem[\protect\citeauthoryear{{Coleman}, {Leleu}, {Alibert}  \&
  {Benz}}{{Coleman} et~al.}{2019}]{Coleman19}
{Coleman} G.~A.~L.,  {Leleu} A.,  {Alibert} Y.,   {Benz} W.,  2019, \mn@doi
  [\aap] {10.1051/0004-6361/201935922}, \href
  {https://ui.adsabs.harvard.edu/abs/2019A&A...631A...7C} {631, A7}

\bibitem[\protect\citeauthoryear{{Cox}}{{Cox}}{2000}]{Cox}
{Cox} A.~N.,  2000, {Allen's astrophysical quantities}

\bibitem[\protect\citeauthoryear{{D'Angelo} \& {Marzari}}{{D'Angelo} \&
  {Marzari}}{2012}]{Dangelo12}
{D'Angelo} G.,  {Marzari} F.,  2012, \mn@doi [\apj]
  {10.1088/0004-637X/757/1/50}, \href
  {http://adsabs.harvard.edu/abs/2012ApJ...757...50D} {757, 50}

\bibitem[\protect\citeauthoryear{{Damasso} et~al.,}{{Damasso}
  et~al.}{2020}]{Damasso20}
{Damasso} M.,  et~al., 2020, \mn@doi [Science Advances]
  {10.1126/sciadv.aax7467}, \href
  {https://ui.adsabs.harvard.edu/abs/2020SciA....6.7467D} {6, eaax7467}

\bibitem[\protect\citeauthoryear{{Delbo}, {Walsh}, {Bolin}, {Avdellidou}  \&
  {Morbidelli}}{{Delbo} et~al.}{2017}]{Delbo17}
{Delbo} M.,  {Walsh} K.,  {Bolin} B.,  {Avdellidou} C.,   {Morbidelli} A.,
  2017, \mn@doi [Science] {10.1126/science.aam6036}, \href
  {https://ui.adsabs.harvard.edu/abs/2017Sci...357.1026D} {357, 1026}

\bibitem[\protect\citeauthoryear{{Desch} \& {Turner}}{{Desch} \&
  {Turner}}{2015}]{DeschTurner2015}
{Desch} S.~J.,  {Turner} N.~J.,  2015, \mn@doi [\apj]
  {10.1088/0004-637X/811/2/156}, \href
  {http://adsabs.harvard.edu/abs/2015ApJ...811..156D} {811, 156}

\bibitem[\protect\citeauthoryear{{Dittrich}, {Klahr}  \& {Johansen}}{{Dittrich}
  et~al.}{2013}]{Dittrich13}
{Dittrich} K.,  {Klahr} H.,   {Johansen} A.,  2013, \mn@doi [\apj]
  {10.1088/0004-637X/763/2/117}, \href
  {http://adsabs.harvard.edu/abs/2013ApJ...763..117D} {763, 117}

\bibitem[\protect\citeauthoryear{{Dr{\k{a}}{\.z}kowska} \&
  {Alibert}}{{Dr{\k{a}}{\.z}kowska} \& {Alibert}}{2017}]{Drazkowska17}
{Dr{\k{a}}{\.z}kowska} J.,  {Alibert} Y.,  2017, \mn@doi [\aap]
  {10.1051/0004-6361/201731491}, \href
  {https://ui.adsabs.harvard.edu/abs/2017A&A...608A..92D} {608, A92}

\bibitem[\protect\citeauthoryear{{Dullemond}, {Hollenbach}, {Kamp}  \&
  {D'Alessio}}{{Dullemond} et~al.}{2007}]{Dullemond}
{Dullemond} C.~P.,  {Hollenbach} D.,  {Kamp} I.,   {D'Alessio} P.,  2007,
  Protostars and Planets V, \href
  {http://adsabs.harvard.edu/abs/2007prpl.conf..555D} {pp 555--572}

\bibitem[\protect\citeauthoryear{{Eriksson}, {Johansen}  \& {Liu}}{{Eriksson}
  et~al.}{2020}]{Eriksson20}
{Eriksson} L. E.~J.,  {Johansen} A.,   {Liu} B.,  2020, \mn@doi [\aap]
  {10.1051/0004-6361/201937037}, \href
  {https://ui.adsabs.harvard.edu/abs/2020A&A...635A.110E} {635, A110}

\bibitem[\protect\citeauthoryear{{Flock}, {Turner}, {Mulders}, {Hasegawa},
  {Nelson}  \& {Bitsch}}{{Flock} et~al.}{2019}]{Flock19}
{Flock} M.,  {Turner} N.~J.,  {Mulders} G.~D.,  {Hasegawa} Y.,  {Nelson} R.~P.,
    {Bitsch} B.,  2019, \mn@doi [\aap] {10.1051/0004-6361/201935806}, \href
  {https://ui.adsabs.harvard.edu/abs/2019A&A...630A.147F} {630, A147}

\bibitem[\protect\citeauthoryear{{Fortier}, {Alibert}, {Carron}, {Benz}  \&
  {Dittkrist}}{{Fortier} et~al.}{2013}]{Fortier13}
{Fortier} A.,  {Alibert} Y.,  {Carron} F.,  {Benz} W.,   {Dittkrist} K.~M.,
  2013, \mn@doi [\aap] {10.1051/0004-6361/201220241}, \href
  {https://ui.adsabs.harvard.edu/abs/2013A&A...549A..44F} {549, A44}

\bibitem[\protect\citeauthoryear{{Fromang} \& {Nelson}}{{Fromang} \&
  {Nelson}}{2006}]{FromangNelson2006}
{Fromang} S.,  {Nelson} R.~P.,  2006, \mn@doi [\aap]
  {10.1051/0004-6361:20065643}, \href
  {http://adsabs.harvard.edu/abs/2006A%26A...457..343F} {457, 343}

\bibitem[\protect\citeauthoryear{{Gillon} et~al.,}{{Gillon}
  et~al.}{2017}]{GillonTrappist17}
{Gillon} M.,  et~al., 2017, \mn@doi [\nat] {10.1038/nature21360}, \href
  {http://adsabs.harvard.edu/abs/2017Natur.542..456G} {542, 456}

\bibitem[\protect\citeauthoryear{{Goldreich} \& {Ward}}{{Goldreich} \&
  {Ward}}{1973}]{GoldreichWard73}
{Goldreich} P.,  {Ward} W.~R.,  1973, \mn@doi [\apj] {10.1086/152291}, \href
  {https://ui.adsabs.harvard.edu/abs/1973ApJ...183.1051G} {183, 1051}

\bibitem[\protect\citeauthoryear{{Gorti} \& {Hollenbach}}{{Gorti} \&
  {Hollenbach}}{2009}]{Gorti09a}
{Gorti} U.,  {Hollenbach} D.,  2009, \mn@doi [\apj]
  {10.1088/0004-637X/690/2/1539}, \href
  {https://ui.adsabs.harvard.edu/abs/2009ApJ...690.1539G} {690, 1539}

\bibitem[\protect\citeauthoryear{{Gorti}, {Dullemond}  \& {Hollenbach}}{{Gorti}
  et~al.}{2009}]{Gorti09}
{Gorti} U.,  {Dullemond} C.~P.,   {Hollenbach} D.,  2009, \mn@doi [\apj]
  {10.1088/0004-637X/705/2/1237}, \href
  {https://ui.adsabs.harvard.edu/abs/2009ApJ...705.1237G} {705, 1237}

\bibitem[\protect\citeauthoryear{{Gorti}, {Hollenbach}  \& {Dullemond}}{{Gorti}
  et~al.}{2015}]{Gorti15}
{Gorti} U.,  {Hollenbach} D.,   {Dullemond} C.~P.,  2015, \mn@doi [\apj]
  {10.1088/0004-637X/804/1/29}, \href
  {https://ui.adsabs.harvard.edu/abs/2015ApJ...804...29G} {804, 29}

\bibitem[\protect\citeauthoryear{{Haisch}, {Lada}  \& {Lada}}{{Haisch}
  et~al.}{2001}]{Haisch01}
{Haisch} Karl~E. J.,  {Lada} E.~A.,   {Lada} C.~J.,  2001, \mn@doi [\apjl]
  {10.1086/320685}, \href
  {https://ui.adsabs.harvard.edu/abs/2001ApJ...553L.153H} {553, L153}

\bibitem[\protect\citeauthoryear{{Haworth}, {Clarke}, {Rahman}, {Winter}  \&
  {Facchini}}{{Haworth} et~al.}{2018}]{Haworth18}
{Haworth} T.~J.,  {Clarke} C.~J.,  {Rahman} W.,  {Winter} A.~J.,   {Facchini}
  S.,  2018, \mn@doi [\mnras] {10.1093/mnras/sty2323}, \href
  {https://ui.adsabs.harvard.edu/abs/2018MNRAS.481..452H} {481, 452}

\bibitem[\protect\citeauthoryear{{Ida}, {Lin}  \& {Nagasawa}}{{Ida}
  et~al.}{2013}]{Ida13}
{Ida} S.,  {Lin} D.~N.~C.,   {Nagasawa} M.,  2013, \mn@doi [\apj]
  {10.1088/0004-637X/775/1/42}, \href
  {http://adsabs.harvard.edu/abs/2013ApJ...775...42I} {775, 42}

\bibitem[\protect\citeauthoryear{{Inaba} \& {Ikoma}}{{Inaba} \&
  {Ikoma}}{2003}]{Inaba}
{Inaba} S.,  {Ikoma} M.,  2003, \mn@doi [\aap] {10.1051/0004-6361:20031248},
  \href {http://adsabs.harvard.edu/abs/2003A%26A...410..711I} {410, 711}

\bibitem[\protect\citeauthoryear{{Inaba}, {Tanaka}, {Nakazawa}, {Wetherill}  \&
  {Kokubo}}{{Inaba} et~al.}{2001}]{Inaba01}
{Inaba} S.,  {Tanaka} H.,  {Nakazawa} K.,  {Wetherill} G.~W.,   {Kokubo} E.,
  2001, \mn@doi [\icarus] {10.1006/icar.2000.6533}, \href
  {https://ui.adsabs.harvard.edu/abs/2001Icar..149..235I} {149, 235}

\bibitem[\protect\citeauthoryear{{Johansen} \& {Lambrechts}}{{Johansen} \&
  {Lambrechts}}{2017}]{Johansen17}
{Johansen} A.,  {Lambrechts} M.,  2017, \mn@doi [Annual Review of Earth and
  Planetary Sciences] {10.1146/annurev-earth-063016-020226}, \href
  {http://adsabs.harvard.edu/abs/2017AREPS..45..359J} {45, 359}

\bibitem[\protect\citeauthoryear{{Johansen}, {Oishi}, {Mac Low}, {Klahr},
  {Henning}  \& {Youdin}}{{Johansen} et~al.}{2007}]{Johansen07}
{Johansen} A.,  {Oishi} J.~S.,  {Mac Low} M.-M.,  {Klahr} H.,  {Henning} T.,
  {Youdin} A.,  2007, \mn@doi [\nat] {10.1038/nature06086}, \href
  {http://adsabs.harvard.edu/abs/2007Natur.448.1022J} {448, 1022}

\bibitem[\protect\citeauthoryear{{Johansen}, {Youdin}  \& {Klahr}}{{Johansen}
  et~al.}{2009a}]{JohansenYoudin2009}
{Johansen} A.,  {Youdin} A.,   {Klahr} H.,  2009a, \mn@doi [\apj]
  {10.1088/0004-637X/697/2/1269}, \href
  {http://adsabs.harvard.edu/abs/2009ApJ...697.1269J} {697, 1269}

\bibitem[\protect\citeauthoryear{{Johansen}, {Youdin}  \& {Mac Low}}{{Johansen}
  et~al.}{2009b}]{Johansen09}
{Johansen} A.,  {Youdin} A.,   {Mac Low} M.-M.,  2009b, \mn@doi [\apjl]
  {10.1088/0004-637X/704/2/L75}, \href
  {https://ui.adsabs.harvard.edu/abs/2009ApJ...704L..75J} {704, L75}

\bibitem[\protect\citeauthoryear{{Johansen}, {Youdin}  \&
  {Lithwick}}{{Johansen} et~al.}{2012}]{Johansen12}
{Johansen} A.,  {Youdin} A.~N.,   {Lithwick} Y.,  2012, \mn@doi [\aap]
  {10.1051/0004-6361/201117701}, \href
  {https://ui.adsabs.harvard.edu/abs/2012A&A...537A.125J} {537, A125}

\bibitem[\protect\citeauthoryear{{Johansen}, {Blum}, {Tanaka}, {Ormel},
  {Bizzarro}  \& {Rickman}}{{Johansen} et~al.}{2014}]{Johansen14}
{Johansen} A.,  {Blum} J.,  {Tanaka} H.,  {Ormel} C.,  {Bizzarro} M.,
  {Rickman} H.,  2014, in {Beuther} H.,  {Klessen} R.~S.,  {Dullemond} C.~P.,
  {Henning} T.,  eds, Protostars and Planets VI. p.~547 (\mn@eprint {arXiv}
  {1402.1344}), \mn@doi{10.2458/azu_uapress_9780816531240-ch024}

\bibitem[\protect\citeauthoryear{{Johansen}, {Mac Low}, {Lacerda}  \&
  {Bizzarro}}{{Johansen} et~al.}{2015}]{Johansen15}
{Johansen} A.,  {Mac Low} M.-M.,  {Lacerda} P.,   {Bizzarro} M.,  2015, \mn@doi
  [Science Advances] {10.1126/sciadv.1500109}, \href
  {https://ui.adsabs.harvard.edu/abs/2015SciA....1E0109J} {1, 1500109}

\bibitem[\protect\citeauthoryear{{Johansen}, {Ronnet}, {Bizzarro}, {Schiller},
  {Lambrechts}, {Nordlund}  \& {Lammer}}{{Johansen} et~al.}{2021}]{Johansen21}
{Johansen} A.,  {Ronnet} T.,  {Bizzarro} M.,  {Schiller} M.,  {Lambrechts} M.,
  {Nordlund} {\r{A}}.,   {Lammer} H.,  2021, \mn@doi [Science Advances]
  {10.1126/sciadv.abc0444}, \href
  {https://ui.adsabs.harvard.edu/abs/2021SciA....7..444J} {7, eabc0444}

\bibitem[\protect\citeauthoryear{{Kokubo} \& {Ida}}{{Kokubo} \&
  {Ida}}{1996}]{Kokubo96}
{Kokubo} E.,  {Ida} S.,  1996, \mn@doi [\icarus] {10.1006/icar.1996.0148},
  \href {https://ui.adsabs.harvard.edu/abs/1996Icar..123..180K} {123, 180}

\bibitem[\protect\citeauthoryear{{Kokubo} \& {Ida}}{{Kokubo} \&
  {Ida}}{1998}]{Kokubo98}
{Kokubo} E.,  {Ida} S.,  1998, \mn@doi [\icarus] {10.1006/icar.1997.5840},
  \href {http://adsabs.harvard.edu/abs/1998Icar..131..171K} {131, 171}

\bibitem[\protect\citeauthoryear{{Lambrechts} \& {Johansen}}{{Lambrechts} \&
  {Johansen}}{2012}]{Lambrechts12}
{Lambrechts} M.,  {Johansen} A.,  2012, \mn@doi [\aap]
  {10.1051/0004-6361/201219127}, \href
  {http://adsabs.harvard.edu/abs/2012A%26A...544A..32L} {544, A32}

\bibitem[\protect\citeauthoryear{{Lambrechts} \& {Johansen}}{{Lambrechts} \&
  {Johansen}}{2014}]{Lambrechts14}
{Lambrechts} M.,  {Johansen} A.,  2014, \mn@doi [\aap]
  {10.1051/0004-6361/201424343}, \href
  {http://adsabs.harvard.edu/abs/2014A%26A...572A.107L} {572, A107}

\bibitem[\protect\citeauthoryear{{Lambrechts}, {Johansen}  \&
  {Morbidelli}}{{Lambrechts} et~al.}{2014}]{Lambrechts14b}
{Lambrechts} M.,  {Johansen} A.,   {Morbidelli} A.,  2014, \mn@doi [\aap]
  {10.1051/0004-6361/201423814}, \href
  {http://adsabs.harvard.edu/abs/2014A%26A...572A..35L} {572, A35}

\bibitem[\protect\citeauthoryear{{Lambrechts}, {Morbidelli}, {Jacobson},
  {Johansen}, {Bitsch}, {Izidoro}  \& {Raymond}}{{Lambrechts}
  et~al.}{2019}]{Lambrechts19}
{Lambrechts} M.,  {Morbidelli} A.,  {Jacobson} S.~A.,  {Johansen} A.,  {Bitsch}
  B.,  {Izidoro} A.,   {Raymond} S.~N.,  2019, \mn@doi [\aap]
  {10.1051/0004-6361/201834229}, \href
  {https://ui.adsabs.harvard.edu/abs/2019A&A...627A..83L} {627, A83}

\bibitem[\protect\citeauthoryear{{Lenz}, {Klahr}  \& {Birnstiel}}{{Lenz}
  et~al.}{2019}]{Lenz19}
{Lenz} C.~T.,  {Klahr} H.,   {Birnstiel} T.,  2019, \mn@doi [\apj]
  {10.3847/1538-4357/ab05d9}, \href
  {https://ui.adsabs.harvard.edu/abs/2019ApJ...874...36L} {874, 36}

\bibitem[\protect\citeauthoryear{{Lenz}, {Klahr}, {Birnstiel}, {Kretke}  \&
  {Stammler}}{{Lenz} et~al.}{2020}]{Lenz20}
{Lenz} C.~T.,  {Klahr} H.,  {Birnstiel} T.,  {Kretke} K.,   {Stammler} S.,
  2020, \mn@doi [\aap] {10.1051/0004-6361/202037878}, \href
  {https://ui.adsabs.harvard.edu/abs/2020A&A...640A..61L} {640, A61}

\bibitem[\protect\citeauthoryear{{Li}, {Youdin}  \& {Simon}}{{Li}
  et~al.}{2019}]{Li19}
{Li} R.,  {Youdin} A.~N.,   {Simon} J.~B.,  2019, \mn@doi [\apj]
  {10.3847/1538-4357/ab480d}, \href
  {https://ui.adsabs.harvard.edu/abs/2019ApJ...885...69L} {885, 69}

\bibitem[\protect\citeauthoryear{{Lin} \& {Papaloizou}}{{Lin} \&
  {Papaloizou}}{1986}]{LinPapaloizou86}
{Lin} D.~N.~C.,  {Papaloizou} J.,  1986, \mn@doi [\apj] {10.1086/164653}, \href
  {http://adsabs.harvard.edu/abs/1986ApJ...309..846L} {309, 846}

\bibitem[\protect\citeauthoryear{{Liu} \& {Ormel}}{{Liu} \&
  {Ormel}}{2018}]{Liu18}
{Liu} B.,  {Ormel} C.~W.,  2018, \mn@doi [\aap] {10.1051/0004-6361/201732307},
  \href {http://adsabs.harvard.edu/abs/2018A%26A...615A.138L} {615, A138}

\bibitem[\protect\citeauthoryear{{Liu}, {Ormel}  \& {Johansen}}{{Liu}
  et~al.}{2019}]{Liu19}
{Liu} B.,  {Ormel} C.~W.,   {Johansen} A.,  2019, \mn@doi [\aap]
  {10.1051/0004-6361/201834174}, \href
  {https://ui.adsabs.harvard.edu/abs/2019A&A...624A.114L} {624, A114}

\bibitem[\protect\citeauthoryear{{Liu}, {Lambrechts}, {Johansen}, {Pascucci}
  \& {Henning}}{{Liu} et~al.}{2020}]{Liu20}
{Liu} B.,  {Lambrechts} M.,  {Johansen} A.,  {Pascucci} I.,   {Henning} T.,
  2020, \mn@doi [\aap] {10.1051/0004-6361/202037720}, \href
  {https://ui.adsabs.harvard.edu/abs/2020A&A...638A..88L} {638, A88}

\bibitem[\protect\citeauthoryear{{Mamajek}}{{Mamajek}}{2009}]{Mamajek09}
{Mamajek} E.~E.,  2009, in {Usuda} T.,  {Tamura} M.,   {Ishii} M.,  eds,
  American Institute of Physics Conference Series Vol. 1158, American Institute
  of Physics Conference Series. pp 3--10 (\mn@eprint {arXiv} {0906.5011}),
  \mn@doi{10.1063/1.3215910}

\bibitem[\protect\citeauthoryear{{Marleau}, {Coleman}, {Leleu}  \&
  {Mordasini}}{{Marleau} et~al.}{2019}]{Marleau19}
{Marleau} G.-D.,  {Coleman} G. A.~L.,  {Leleu} A.,   {Mordasini} C.,  2019,
  \mn@doi [\aap] {10.1051/0004-6361/201833597}, \href
  {https://ui.adsabs.harvard.edu/abs/2019A&A...624A..20M} {624, A20}

\bibitem[\protect\citeauthoryear{{Matsuyama}, {Johnstone}  \&
  {Hartmann}}{{Matsuyama} et~al.}{2003}]{Matsuyama03}
{Matsuyama} I.,  {Johnstone} D.,   {Hartmann} L.,  2003, \mn@doi [\apj]
  {10.1086/344638}, \href
  {https://ui.adsabs.harvard.edu/abs/2003ApJ...582..893M} {582, 893}

\bibitem[\protect\citeauthoryear{{Mayor} \& {Queloz}}{{Mayor} \&
  {Queloz}}{1995}]{MayorQueloz}
{Mayor} M.,  {Queloz} D.,  1995, \mn@doi [\nat] {10.1038/378355a0}, \href
  {http://adsabs.harvard.edu/abs/1995Natur.378..355M} {378, 355}

\bibitem[\protect\citeauthoryear{{McNally}, {Nelson}  \&
  {Paardekooper}}{{McNally} et~al.}{2018}]{McNally18}
{McNally} C.~P.,  {Nelson} R.~P.,   {Paardekooper} S.-J.,  2018, \mn@doi
  [\mnras] {10.1093/mnras/sty905}, \href
  {http://adsabs.harvard.edu/abs/2018MNRAS.477.4596M} {477, 4596}

\bibitem[\protect\citeauthoryear{Meerschaert, Roy  \& Shao}{Meerschaert
  et~al.}{2012}]{Meerschaert12}
Meerschaert M.~M.,  Roy P.,   Shao Q.,  2012, \mn@doi [Communications in
  Statistics - Theory and Methods] {10.1080/03610926.2011.552828}, 41, 1839

\bibitem[\protect\citeauthoryear{{Menou} \& {Goodman}}{{Menou} \&
  {Goodman}}{2004}]{Menou}
{Menou} K.,  {Goodman} J.,  2004, \mn@doi [\apj] {10.1086/382947}, \href
  {http://adsabs.harvard.edu/abs/2004ApJ...606..520M} {606, 520}

\bibitem[\protect\citeauthoryear{{Mihalas} \& {Mihalas}}{{Mihalas} \&
  {Mihalas}}{1984}]{Mihalas}
{Mihalas} D.,  {Mihalas} B.~W.,  1984, {Foundations of radiation hydrodynamics}

\bibitem[\protect\citeauthoryear{{Morbidelli}, {Bottke}, {Nesvorn{\'y}}  \&
  {Levison}}{{Morbidelli} et~al.}{2009}]{Morbidelli09}
{Morbidelli} A.,  {Bottke} W.~F.,  {Nesvorn{\'y}} D.,   {Levison} H.~F.,  2009,
  \mn@doi [\icarus] {10.1016/j.icarus.2009.07.011}, \href
  {https://ui.adsabs.harvard.edu/abs/2009Icar..204..558M} {204, 558}

\bibitem[\protect\citeauthoryear{{Mordasini}, {Alibert}  \& {Benz}}{{Mordasini}
  et~al.}{2009}]{Mords09}
{Mordasini} C.,  {Alibert} Y.,   {Benz} W.,  2009, \mn@doi [\aap]
  {10.1051/0004-6361/200810301}, \href
  {http://adsabs.harvard.edu/abs/2009A%26A...501.1139M} {501, 1139}

\bibitem[\protect\citeauthoryear{{Mordasini}, {Molli{\`e}re}, {Dittkrist},
  {Jin}  \& {Alibert}}{{Mordasini} et~al.}{2015}]{Mordasini15}
{Mordasini} C.,  {Molli{\`e}re} P.,  {Dittkrist} K.-M.,  {Jin} S.,   {Alibert}
  Y.,  2015, \mn@doi [International Journal of Astrobiology]
  {10.1017/S1473550414000263}, \href
  {http://adsabs.harvard.edu/abs/2015IJAsB..14..201M} {14, 201}

\bibitem[\protect\citeauthoryear{{Nakagawa}, {Sekiya}  \& {Hayashi}}{{Nakagawa}
  et~al.}{1986}]{Nakagawa86}
{Nakagawa} Y.,  {Sekiya} M.,   {Hayashi} C.,  1986, \mn@doi [\icarus]
  {10.1016/0019-1035(86)90121-1}, \href
  {https://ui.adsabs.harvard.edu/abs/1986Icar...67..375N} {67, 375}

\bibitem[\protect\citeauthoryear{{Ohtsuki}}{{Ohtsuki}}{1999}]{Ohtsuki99}
{Ohtsuki} K.,  1999, \mn@doi [\icarus] {10.1006/icar.1998.6041}, \href
  {https://ui.adsabs.harvard.edu/abs/1999Icar..137..152O} {137, 152}

\bibitem[\protect\citeauthoryear{{Ormel} \& {Klahr}}{{Ormel} \&
  {Klahr}}{2010}]{OrmelKlahr2010}
{Ormel} C.~W.,  {Klahr} H.~H.,  2010, \mn@doi [\aap]
  {10.1051/0004-6361/201014903}, \href
  {http://adsabs.harvard.edu/abs/2010A%26A...520A..43O} {520, A43}

\bibitem[\protect\citeauthoryear{{Ormel} \& {Liu}}{{Ormel} \&
  {Liu}}{2018}]{Ormel18}
{Ormel} C.~W.,  {Liu} B.,  2018, \mn@doi [\aap] {10.1051/0004-6361/201732562},
  \href {http://adsabs.harvard.edu/abs/2018A%26A...615A.178O} {615, A178}

\bibitem[\protect\citeauthoryear{{Owen}, {Clarke}  \& {Ercolano}}{{Owen}
  et~al.}{2012}]{Owen12}
{Owen} J.~E.,  {Clarke} C.~J.,   {Ercolano} B.,  2012, \mn@doi [\mnras]
  {10.1111/j.1365-2966.2011.20337.x}, \href
  {https://ui.adsabs.harvard.edu/abs/2012MNRAS.422.1880O} {422, 1880}

\bibitem[\protect\citeauthoryear{{Paardekooper}}{{Paardekooper}}{2014}]{Paardekooper2014}
{Paardekooper} S.-J.,  2014, \mn@doi [\mnras] {10.1093/mnras/stu1542}, \href
  {http://adsabs.harvard.edu/abs/2014MNRAS.444.2031P} {444, 2031}

\bibitem[\protect\citeauthoryear{{Paardekooper} \& {Mellema}}{{Paardekooper} \&
  {Mellema}}{2006}]{PaardekooperMellema06}
{Paardekooper} S.-J.,  {Mellema} G.,  2006, \mn@doi [\aap]
  {10.1051/0004-6361:20066304}, \href
  {http://adsabs.harvard.edu/abs/2006A%26A...459L..17P} {459, L17}

\bibitem[\protect\citeauthoryear{{Paardekooper}, {Baruteau}, {Crida}  \&
  {Kley}}{{Paardekooper} et~al.}{2010}]{pdk10}
{Paardekooper} S.-J.,  {Baruteau} C.,  {Crida} A.,   {Kley} W.,  2010, \mn@doi
  [\mnras] {10.1111/j.1365-2966.2009.15782.x}, \href
  {http://adsabs.harvard.edu/abs/2010MNRAS.401.1950P} {401, 1950}

\bibitem[\protect\citeauthoryear{{Paardekooper}, {Baruteau}  \&
  {Kley}}{{Paardekooper} et~al.}{2011}]{pdk11}
{Paardekooper} S.-J.,  {Baruteau} C.,   {Kley} W.,  2011, \mn@doi [\mnras]
  {10.1111/j.1365-2966.2010.17442.x}, \href
  {http://adsabs.harvard.edu/abs/2011MNRAS.410..293P} {410, 293}

\bibitem[\protect\citeauthoryear{{Papaloizou} \& {Nelson}}{{Papaloizou} \&
  {Nelson}}{2003}]{PapaloizouNelson2003}
{Papaloizou} J.~C.~B.,  {Nelson} R.~P.,  2003, \mn@doi [\mnras]
  {10.1046/j.1365-8711.2003.06246.x}, \href
  {http://adsabs.harvard.edu/abs/2003MNRAS.339..983P} {339, 983}

\bibitem[\protect\citeauthoryear{{Pollack}, {Hubickyj}, {Bodenheimer},
  {Lissauer}, {Podolak}  \& {Greenzweig}}{{Pollack} et~al.}{1996}]{Pollack}
{Pollack} J.~B.,  {Hubickyj} O.,  {Bodenheimer} P.,  {Lissauer} J.~J.,
  {Podolak} M.,   {Greenzweig} Y.,  1996, \mn@doi [\icarus]
  {10.1006/icar.1996.0190}, \href
  {http://adsabs.harvard.edu/abs/1996Icar..124...62P} {124, 62}

\bibitem[\protect\citeauthoryear{{Poon}, {Nelson}  \& {Coleman}}{{Poon}
  et~al.}{2021}]{Poon21}
{Poon} S. T.~S.,  {Nelson} R.~P.,   {Coleman} G. A.~L.,  2021, \mn@doi [\mnras]
  {10.1093/mnras/stab1466}, \href
  {https://ui.adsabs.harvard.edu/abs/2021MNRAS.tmp.1402P} {}

\bibitem[\protect\citeauthoryear{{Pu} \& {Wu}}{{Pu} \& {Wu}}{2015}]{PuWu2015}
{Pu} B.,  {Wu} Y.,  2015, \mn@doi [\apj] {10.1088/0004-637X/807/1/44}, \href
  {http://adsabs.harvard.edu/abs/2015ApJ...807...44P} {807, 44}

\bibitem[\protect\citeauthoryear{{Rafikov}}{{Rafikov}}{2004}]{Rafikov04}
{Rafikov} R.~R.,  2004, \mn@doi [\aj] {10.1086/423216}, \href
  {https://ui.adsabs.harvard.edu/abs/2004AJ....128.1348R} {128, 1348}

\bibitem[\protect\citeauthoryear{{Raymond}, {Kokubo}, {Morbidelli}, {Morishima}
   \& {Walsh}}{{Raymond} et~al.}{2014}]{Raymond14}
{Raymond} S.~N.,  {Kokubo} E.,  {Morbidelli} A.,  {Morishima} R.,   {Walsh}
  K.~J.,  2014, in {Beuther} H.,  {Klessen} R.~S.,  {Dullemond} C.~P.,
  {Henning} T.,  eds, Protostars and Planets VI. p.~595 (\mn@eprint {arXiv}
  {1312.1689}), \mn@doi{10.2458/azu_uapress_9780816531240-ch026}

\bibitem[\protect\citeauthoryear{{Ribas}, {Mer{\'\i}n}, {Bouy}  \&
  {Maud}}{{Ribas} et~al.}{2014}]{Ribas14}
{Ribas} {\'A}.,  {Mer{\'\i}n} B.,  {Bouy} H.,   {Maud} L.~T.,  2014, \mn@doi
  [\aap] {10.1051/0004-6361/201322597}, \href
  {https://ui.adsabs.harvard.edu/abs/2014A&A...561A..54R} {561, A54}

\bibitem[\protect\citeauthoryear{{Rice}, {Armitage}, {Wood}  \&
  {Lodato}}{{Rice} et~al.}{2006}]{Rice06}
{Rice} W.~K.~M.,  {Armitage} P.~J.,  {Wood} K.,   {Lodato} G.,  2006, \mn@doi
  [\mnras] {10.1111/j.1365-2966.2006.11113.x}, \href
  {https://ui.adsabs.harvard.edu/abs/2006MNRAS.373.1619R} {373, 1619}

\bibitem[\protect\citeauthoryear{{Robertson} et~al.,}{{Robertson}
  et~al.}{2012}]{Robertson12}
{Robertson} P.,  et~al., 2012, \mn@doi [\apj] {10.1088/0004-637X/749/1/39},
  \href {https://ui.adsabs.harvard.edu/abs/2012ApJ...749...39R} {749, 39}

\bibitem[\protect\citeauthoryear{{Sch{\"a}fer}, {Yang}  \&
  {Johansen}}{{Sch{\"a}fer} et~al.}{2017}]{Schafer17}
{Sch{\"a}fer} U.,  {Yang} C.-C.,   {Johansen} A.,  2017, \mn@doi [\aap]
  {10.1051/0004-6361/201629561}, \href
  {https://ui.adsabs.harvard.edu/abs/2017A&A...597A..69S} {597, A69}

\bibitem[\protect\citeauthoryear{{Schreiber} \& {Klahr}}{{Schreiber} \&
  {Klahr}}{2018}]{Schreiber18}
{Schreiber} A.,  {Klahr} H.,  2018, \mn@doi [\apj] {10.3847/1538-4357/aac3d4},
  \href {https://ui.adsabs.harvard.edu/abs/2018ApJ...861...47S} {861, 47}

\bibitem[\protect\citeauthoryear{{Sellek}, {Booth}  \& {Clarke}}{{Sellek}
  et~al.}{2020}]{Sellek20}
{Sellek} A.~D.,  {Booth} R.~A.,   {Clarke} C.~J.,  2020, \mn@doi [\mnras]
  {10.1093/mnras/stz3528}, \href
  {https://ui.adsabs.harvard.edu/abs/2020MNRAS.492.1279S} {492, 1279}

\bibitem[\protect\citeauthoryear{{Shakura} \& {Sunyaev}}{{Shakura} \&
  {Sunyaev}}{1973}]{Shak}
{Shakura} N.~I.,  {Sunyaev} R.~A.,  1973, \aap, \href
  {http://adsabs.harvard.edu/abs/1973A%26A....24..337S} {24, 337}

\bibitem[\protect\citeauthoryear{{Simon}, {Beckwith}  \& {Armitage}}{{Simon}
  et~al.}{2012}]{Simon12}
{Simon} J.~B.,  {Beckwith} K.,   {Armitage} P.~J.,  2012, \mn@doi [\mnras]
  {10.1111/j.1365-2966.2012.20835.x}, \href
  {http://adsabs.harvard.edu/abs/2012MNRAS.422.2685S} {422, 2685}

\bibitem[\protect\citeauthoryear{{Simon}, {Armitage}, {Li}  \&
  {Youdin}}{{Simon} et~al.}{2016}]{Simon16}
{Simon} J.~B.,  {Armitage} P.~J.,  {Li} R.,   {Youdin} A.~N.,  2016, \mn@doi
  [\apj] {10.3847/0004-637X/822/1/55}, \href
  {https://ui.adsabs.harvard.edu/abs/2016ApJ...822...55S} {822, 55}

\bibitem[\protect\citeauthoryear{{Simon}, {Armitage}, {Youdin}  \&
  {Li}}{{Simon} et~al.}{2017}]{Simon17}
{Simon} J.~B.,  {Armitage} P.~J.,  {Youdin} A.~N.,   {Li} R.,  2017, \mn@doi
  [\apjl] {10.3847/2041-8213/aa8c79}, \href
  {https://ui.adsabs.harvard.edu/abs/2017ApJ...847L..12S} {847, L12}

\bibitem[\protect\citeauthoryear{{Steinacker} \& {Papaloizou}}{{Steinacker} \&
  {Papaloizou}}{2002}]{SteinackerPapaloizou2002}
{Steinacker} A.,  {Papaloizou} J.~C.~B.,  2002, \mn@doi [\apj]
  {10.1086/339892}, \href {http://adsabs.harvard.edu/abs/2002ApJ...571..413S}
  {571, 413}

\bibitem[\protect\citeauthoryear{{Umebayashi} \& {Nakano}}{{Umebayashi} \&
  {Nakano}}{1988}]{UmebayashiNakano1988}
{Umebayashi} T.,  {Nakano} T.,  1988, \mn@doi [Progress of Theoretical Physics
  Supplement] {10.1143/PTPS.96.151}, \href
  {http://adsabs.harvard.edu/abs/1988PThPS..96..151U} {96, 151}

\bibitem[\protect\citeauthoryear{{Voelkel}, {Klahr}, {Mordasini}, {Emsenhuber}
  \& {Lenz}}{{Voelkel} et~al.}{2020}]{Voelkel20}
{Voelkel} O.,  {Klahr} H.,  {Mordasini} C.,  {Emsenhuber} A.,   {Lenz} C.,
  2020, \mn@doi [\aap] {10.1051/0004-6361/202038085}, \href
  {https://ui.adsabs.harvard.edu/abs/2020A&A...642A..75V} {642, A75}

\bibitem[\protect\citeauthoryear{{Weidenschilling}}{{Weidenschilling}}{1977}]{Weidenschilling_77}
{Weidenschilling} S.~J.,  1977, \mn@doi [\mnras] {10.1093/mnras/180.1.57},
  \href {http://adsabs.harvard.edu/abs/1977MNRAS.180...57W} {180, 57}

\bibitem[\protect\citeauthoryear{{Wetherill} \& {Stewart}}{{Wetherill} \&
  {Stewart}}{1989}]{Wetherill89}
{Wetherill} G.~W.,  {Stewart} G.~R.,  1989, \mn@doi [\icarus]
  {10.1016/0019-1035(89)90093-6}, \href
  {https://ui.adsabs.harvard.edu/abs/1989Icar...77..330W} {77, 330}

\bibitem[\protect\citeauthoryear{{Wetherill} \& {Stewart}}{{Wetherill} \&
  {Stewart}}{1993}]{Wetherill93}
{Wetherill} G.~W.,  {Stewart} G.~R.,  1993, \mn@doi [\icarus]
  {10.1006/icar.1993.1166}, \href
  {https://ui.adsabs.harvard.edu/abs/1993Icar..106..190W} {106, 190}

\bibitem[\protect\citeauthoryear{{Winn} \& {Fabrycky}}{{Winn} \&
  {Fabrycky}}{2015}]{Winn15}
{Winn} J.~N.,  {Fabrycky} D.~C.,  2015, \mn@doi [\araa]
  {10.1146/annurev-astro-082214-122246}, \href
  {https://ui.adsabs.harvard.edu/abs/2015ARA&A..53..409W} {53, 409}

\bibitem[\protect\citeauthoryear{{Yang}, {Johansen}  \& {Carrera}}{{Yang}
  et~al.}{2017}]{Yang17}
{Yang} C.~C.,  {Johansen} A.,   {Carrera} D.,  2017, \mn@doi [\aap]
  {10.1051/0004-6361/201630106}, \href
  {https://ui.adsabs.harvard.edu/abs/2017A&A...606A..80Y} {606, A80}

\bibitem[\protect\citeauthoryear{{Youdin} \& {Goodman}}{{Youdin} \&
  {Goodman}}{2005}]{Youdin05}
{Youdin} A.~N.,  {Goodman} J.,  2005, \mn@doi [\apj] {10.1086/426895}, \href
  {https://ui.adsabs.harvard.edu/abs/2005ApJ...620..459Y} {620, 459}

\bibitem[\protect\citeauthoryear{{Youdin} \& {Lithwick}}{{Youdin} \&
  {Lithwick}}{2007}]{Youdin07}
{Youdin} A.~N.,  {Lithwick} Y.,  2007, \mn@doi [\icarus]
  {10.1016/j.icarus.2007.07.012}, \href
  {https://ui.adsabs.harvard.edu/abs/2007Icar..192..588Y} {192, 588}

\bibitem[\protect\citeauthoryear{{Zhu}, {Stone}, {Rafikov}  \& {Bai}}{{Zhu}
  et~al.}{2014}]{ZhuStoneBai2014}
{Zhu} Z.,  {Stone} J.~M.,  {Rafikov} R.~R.,   {Bai} X.-n.,  2014, \mn@doi
  [\apj] {10.1088/0004-637X/785/2/122}, \href
  {http://adsabs.harvard.edu/abs/2014ApJ...785..122Z} {785, 122}

\makeatother
\end{thebibliography}
\label{lastpage}
\end{document}